\documentclass[a4paper,twocolumn,11pt,accepted=2020-09-01]{quantumarticle}
\pdfoutput=1
\usepackage[utf8]{inputenc}
\usepackage[english]{babel}
\usepackage[T1]{fontenc}
\usepackage{amsmath}
\usepackage{hyperref}

\usepackage{tikz}
\usepackage{lipsum}

\usepackage[sort&compress,numbers]{natbib}
\newcommand{\rrangle}{\rangle\!\rangle}
\newcommand{\llangle}{\langle\!\langle}

\begin{document}

\title{Efficient classical simulation of noisy random quantum circuits in one dimension}

\author{Kyungjoo Noh}
\affiliation{Department of Physics, Yale University, New Haven, Connecticut 06520, USA}
\affiliation{AWS Center for Quantum Computing, Pasadena, CA, 91125, USA}
\email{nkyungjo@amazon.com}
\orcid{0000-0002-6318-8472}
\thanks{This work was done before K.N. joined AWS Center for Quantum Computing.}
\author{Liang Jiang}
\affiliation{Pritzker School of Molecular Engineering, University of Chicago, Chicago, Illinois 60637, USA}
\email{liang.jiang@uchicago.edu}
\orcid{0000-0002-0000-9342}
\author{Bill Fefferman}
\affiliation{Department of Computer Science, University of Chicago, Chicago, Illinois 60637, USA}
\email{wjf@uchicago.edu}
\orcid{0000-0002-9627-0210}
\maketitle

\begin{abstract}
Understanding the computational power of noisy intermediate-scale quantum (NISQ) devices is of both fundamental and practical importance to quantum information science. Here, we address the question of whether error-uncorrected noisy quantum computers can provide computational advantage over classical computers. Specifically, we study noisy random circuit sampling in one dimension (or 1D noisy RCS) as a simple model for exploring the effects of noise on the computational power of a noisy quantum device. In particular, we simulate the real-time dynamics of 1D noisy random quantum circuits via matrix product operators (MPOs) and characterize the computational power of the 1D noisy quantum system by using a metric we call MPO entanglement entropy. The latter metric is chosen because it determines the cost of classical MPO simulation. We numerically demonstrate that for the two-qubit gate error rates we considered, there exists a characteristic system size above which adding more qubits does not bring about an exponential growth of the cost of classical MPO simulation of 1D noisy systems. Specifically, we show that above the characteristic system size, there is an optimal circuit depth, independent of the system size, where the MPO entanglement entropy is maximized. Most importantly, the maximum achievable MPO entanglement entropy is bounded by a constant that depends only on the gate error rate, not on the system size. We also provide a heuristic analysis to get the scaling of the maximum achievable MPO entanglement entropy as a function of the gate error rate. The obtained scaling suggests that although the cost of MPO simulation does not increase exponentially in the system size above a certain characteristic system size, it does increase exponentially as the gate error rate decreases, possibly making classical simulation practically not feasible even with state-of-the-art supercomputers. 
\end{abstract}

\section{Introduction}
\label{section:Introduction}

Quantum computers can provide significant computational advantage over classical computers as they can efficiently solve certain important problems that are believed to be not solvable in polynomial time with classical computers. Examples of such problems include integer factorization \cite{Shor1994} and simulation of the real-time dynamics of large quantum systems \cite{Lloyd1996}. While currently available quantum devices are not large and reliable enough to factor a large integer or simulate the dynamics of a large quantum system, it has been established over the past two decades that fault-tolerant quantum computing is in principle possible via quantum error correction \cite{Shor1996,Gottesman2000,Bravyi2005,Knill2005,Gottesman2009,Fowler2012}. However, despite the recent progress in reducing high resource overhead associated with the use of fault-tolerant quantum computing schemes \cite{Fowler2012a,Bravyi2012,Horsman2012,Haah2017,Chamberland2018a,
Chao2018,Chao2018b,Chamberland2019,Litinski2019,Chao2019,
Chamberland2020,Chamberland2020a,Das2020,Delfosse2020,Delfosse2020a,
Chamberland2020b}, large-scale and fault-tolerant quantum computing is not yet within reach of near-term quantum technologies.  

Due to the lack of fault-tolerance, currently available noisy intermediate-scale quantum (NISQ) \cite{Preskill2018} devices are clearly not capable of realizing the full potential of quantum computing. Nevertheless, NISQ devices may be able to provide computational advantage over the best available classical computer in tackling certain computational tasks, whether or not solving them is practically useful. Various proposals for demonstrating such quantum computational advantage with NISQ devices have focused on sampling problems such as IQP \cite{Bremner2011}, boson sampling \cite{Aaronson2011,Hamilton2017}, Fourier sampling \cite{Fefferman2016}, and random circuit sampling (RCS) \cite{Boixo2018}. In particular, various complexity-theoretic hardness results for these sampling problems have made them an appealing proposal for demonstrating quantum computational advantage.

Among these proposals, boson sampling was the first whose hardness was proven to be robust against adversarial total variation distance noise, under reasonable hardness assumptions from computational complexity theory \cite{Aaronson2011}. Motivated by such a rigorous hardness result, experimental realizations of boson sampling followed shortly thereafter \cite{Broome2013,Spring2013,Tillmann2013,Crespi2013}. However, all the previous boson sampling experiments have been performed with a limited number of photons that is not large enough to make the system classically intractable. Moreover, various high-performing classical algorithms that are tailored to boson sampling have been developed \cite{Neville2017,Clifford2018,Renema2018,GarciaPatron2019}. These recent developments have thus made it much more challenging to demonstrate quantum computational advantage via boson sampling.


Over the past few years, RCS has risen as a promising candidate for achieving quantum computational advantage since it can be realized at scale in superconducting qubit systems \cite{Boixo2018}. While initially motivated by the experimental viability, RCS was also recently shown to have similar asymptotic
hardness guarantees as boson sampling \cite{Bouland2019} and complementary hardness evidence was shown in Ref.\ \cite{Aaronson2016}, making RCS an even more compelling proposal. Notably, RCS was implemented in a superconducting qubit system which astonishingly consists of $53$ qubits that are connected via two-qubit gates with very low gate error rates ($p \sim 0.006$) in a planar architecture \cite{Arute2019}. In particular, it was claimed in Ref.\ \cite{Arute2019} that it would take about $10000$ years for a state-of-the-art classical supercomputer to achieve a computational task that is equivalent to the one that their superconducting quantum device has achieved. In contrast, a recent work \cite{Pednault2019} has suggested that a refined simulation technique can bring down the required computing time to just a few days. In any case, what has become clear is that currently available superconducting qubit systems can tackle certain computational tasks that lie close to the borderline of what is achievable and not achievable with classical computing technologies.

Going forward, an important thing to keep in mind is that currently available quantum devices are noisy. Thus, a crucial related question is how the classical computing time needed to simulate such noisy random quantum circuits would scale as a function of the system size and the gate error rate. Thanks to the rigorous complexity-theoretic results \cite{Bremner2011,Aaronson2011,Bouland2019,Movassagh2018,Movassagh2019}, it has been established that in the noiseless case, simulating the outputs of a random quantum circuit cannot be done classically in polynomial time in the system size. Moreover, for boson sampling and RCS, the classical intractability was shown to persist even in the presence of the total variation distance noise under suitable hardness assumptions \cite{Aaronson2011,Bouland2019,Movassagh2018,Movassagh2019}. 

However, modeling noise using only closeness in total variation distance does not suffice to address practically relevant settings such as the setting in which each gate is corrupted by an error channel with a non-zero gate error rate. This is because in realistic settings, the effects of noise dominate in the large circuit depth limit and the system eventually converges to a depolarized state. Thus, it is not immediately clear how much computational power can be gained by adding more qubits to noisy quantum systems. Addressing this question is thus essential for understanding the utility of near-term applications of NISQ technologies. 

\begin{figure}[t!]
\centering
\includegraphics[width=0.48\textwidth]{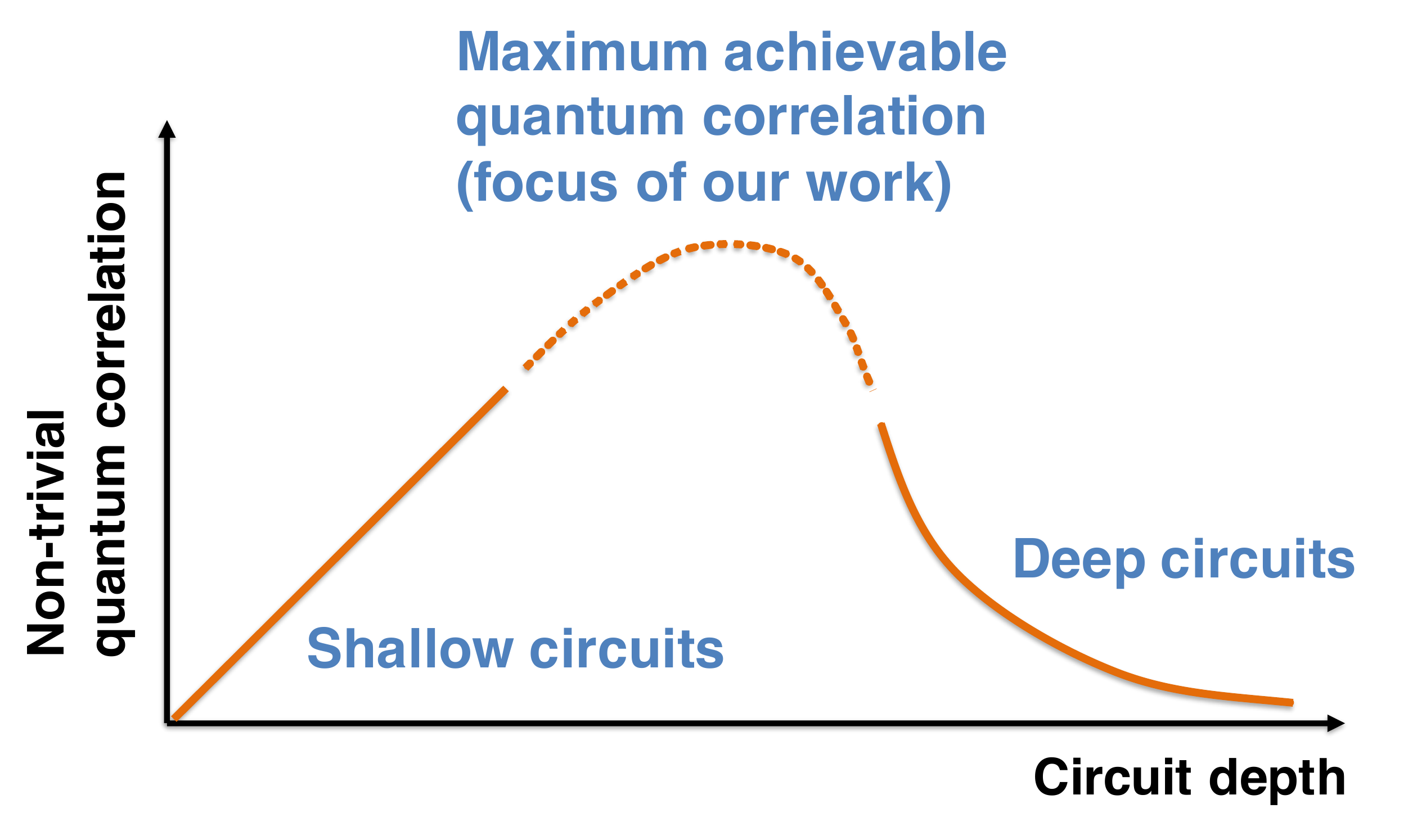}
\caption{Schematic plot of the degree of non-trivial quantum correlation as a function of the circuit depth. When the circuit depth is small, quantum correlation grows linearly in the circuit depth. On the other hand, when the circuit depth is large, the system converges to a depolarized state and thus the non-trivial quantum correlations are washed away. The focus of our work is to understand the optimal regime where the maximum non-trivial quantum correlation is achieved. See also Figs.\ \ref{fig:MPOEE_p_15_10_06} and \ref{fig:MPOEE_n_16_various_p}. }
\label{fig:Schematic_big_picture}
\end{figure}

In this paper, we study noisy random circuit sampling in one dimension (i.e., 1D noisy RCS) as a simple model for exploring the effects of noise on the computational power of a noisy quantum device. Note that since noisy systems eventually converge to a depolarized state, any non-trivial quantum correlations will be washed away in the large circuit depth limit, making the outputs of the system well approximated by a trivial uniform distribution. On the other hand, shallow circuits with a constant circuit depth can also be simulated efficiently via matrix product states (MPSs) \cite{Vidal2003} due to the limited growth of entanglement. Connecting these two extreme cases, we can expect that the degree of non-trivial quantum correlation will be peaked at a certain optimal circuit depth as illustrated in Fig.\ \ref{fig:Schematic_big_picture}. The focus of our work is to understand this optimal regime where the maximum non-trivial quantum correlation is attained. In particular, we explore how hard it is to simulate the 1D noisy system at the optimal circuit depth.     

Note that if one's goal is to approximately simulate ideal random quantum circuits with any non-zero fidelity, it may suffice to use MPSs with a constant bond dimension even for deep circuits (see, e.g., Ref.\ \cite{Zhou2020} and also the discussion in Section \ref{section:Relation to previous results}). However, this is not our goal here and we instead aim to simulate noisy random quantum circuits to any desired accuracy. More specifically, we directly simulate the mixed state dynamics of 1D noisy random quantum circuits by using matrix product operators (MPOs) \cite{Verstraete2004,Zwolak2004}. Note that this is a strictly more challenging task than sampling since any output probability (including marginal and conditional probabilities) can be computed efficiently from an MPO and thus sampling can be done straightforwardly. 

The main contribution of our work is to characterize the computational power of 1D noisy quantum devices by using a metric we call MPO entanglement entropy. We choose the latter metric because it determines the cost of classical MPO simulation as well as the degree of non-trivial quantum correlation of a mixed state. We numerically demonstrate the maximum achievable MPO entanglement entropy is bounded by a constant that depends only on the gate error rate, not on the system size. In other words, the maximum achievable MPO entanglement entropy is saturated at a certain characteristic system size and consequently the required MPO bond dimension does not increase exponentially in the system size above the characteristic system size. Thus, our results indicate that there exists a characteristic system size above which adding more qubits does not help increasing the cost of classically simulating a 1D noisy quantum device in an exponential way. We also provide a heuristic argument to get the scaling of the maximum achievable MPO entanglement entropy as a function of the gate error rate. The obtained scaling suggests that the cost of MPO simulation increases exponentially as the gate error rate decreases, possibly making classical simulation practically not feasible even with a state-of-the-art supercomputer.

\begin{figure*}[t!]
\centering
\includegraphics[width=0.9\textwidth]{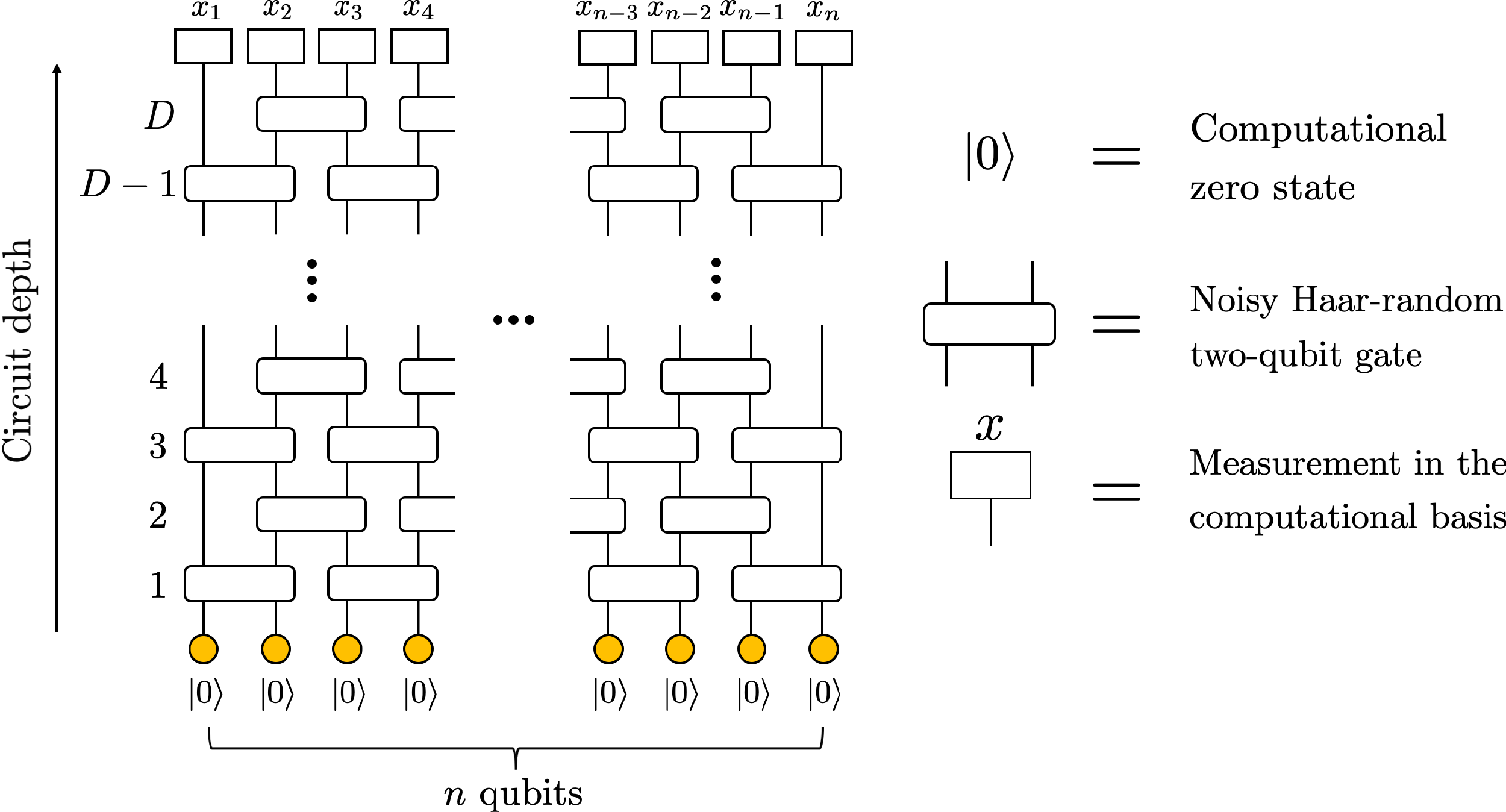}
\caption{Noisy random circuit sampling in one dimension. Each noisy two-qubit gate is given by a $4\times 4$ Haar-random unitary operation followed by a two-qubit depolarization channel $\mathcal{N}_{2}[p]$ with an error rate $p$. At the end of the circuit, all the qubits are measured in the computational basis. For simplicity, we only consider even number of qubits. Although the maximum circuit depth $D$ is chosen to be even in the schematic illustration, we allow $D$ to be odd as well. }
\label{fig:NoisyRCS1D}
\end{figure*}

Our paper is organized as follows. In Section \ref{section:Problem setup}, we formulate the problem of 1D noisy RCS. In Section \ref{section:Matrix product operators}, we briefly review matrix product operators (MPOs) and define MPO entanglement entropy that determines the cost of classical MPO simulation. In Section \ref{section:Main results}, we present the main numerical results on the MPO simulation of 1D noisy RCS. Moreover, in Subsection \ref{subsection:Maximum MPO entanglement entropy}, we provide a heuristic scaling analysis of the maximum achievable MPO entanglement entropy as a function of the gate error rate. In Section \ref{section:Relation to previous results}, we discuss the relation of our results to previous results. We conclude the paper by outlining several open questions in Section \ref{section:Summary and open questions}.

\section{Problem setup}
\label{section:Problem setup}

In this section, we formulate the problem of noisy random circuit sampling in one dimension (i.e., 1D noisy RCS). We also introduce two-qubit depolarization error model which we assume to get the numerical results in Section \ref{section:Main results}. 

\subsection{Noisy random circuit sampling in one dimension}
\label{subsection:Noisy random circuit sampling in one dimension}

Consider $n$ qubits laid out in a one-dimensional chain as shown in Fig.\ \ref{fig:NoisyRCS1D}. For simplicity, we only consider even $n$. Initially, all the qubits are prepared in the computational zero state, i.e.,
\begin{align}
|\psi_{0}\rangle &= |\vec{0}\rangle \equiv |0\rangle^{\otimes n}.  
\end{align}
In odd (or even) time steps, Haar-random two-qubit gates are applied to the $l^{\textrm{th}}$ and the $l+1^{\textrm{th}}$ qubits for $l\in\lbrace 1,3,\cdots,n-1 \rbrace$ (or $l\in\lbrace 2,4,\cdots,n-2 \rbrace$). Eventually, we will assume that each Haar-random two-qubit gate is corrupted by a noisy two-qubit CPTP map \cite{Choi1975} acting on the same sites. However, we assume that two-qubit gates are noiseless for now. The circuit depth $D$ is defined as the number of time steps. Although only the case with an even $D$ is shown in Fig.\ \ref{fig:NoisyRCS1D}, $D$ can also be odd. At the end of the circuit, all the qubits are measured in the computational basis $\lbrace |0\rangle,|1\rangle \rbrace$. Thus, we are left with an output $n$-bit string 
\begin{align}
\vec{x} = (x_{1},\cdots, x_{n}) \in \lbrace 0, 1\rbrace^{n}, 
\end{align}
which is drawn from a probability distribution 
\begin{align}
P_{\mathcal{C}}(\vec{x}) \equiv |\langle \vec{x}| \hat{U}_{\mathcal{C}}|\vec{0}\rangle|^{2}. 
\end{align}
Here, $\hat{U}_{\mathcal{C}}$ is the unitary operator associated with an instance of a depth-$D$ random circuit $\mathcal{C}$. Exact RCS is a sampling problem where the goal is to sample exactly from the ideal output distribution $P_{\mathcal{C}}$ of a given quantum circuit $\mathcal{C}$. One can also define an approximate version of RCS, i.e., approximate RCS, which is a sampling problem where the goal is to sample from a distribution $P'_{\mathcal{C}}$ that is $\epsilon$-close to the ideal output distribution $P_{\mathcal{C}})$ in total variation distance, i.e., 
\begin{align}
|\!| P'_{\mathcal{C}}- P_{\mathcal{C}} |\!| \equiv \frac{1}{2}\sum_{\vec{x} \in \lbrace 0,1 \rbrace^{n} } | P'_{\mathcal{C}}(\vec{x}) - P_{\mathcal{C}}(\vec{x}) | \le \epsilon. \label{eq:total variation distance noise}
\end{align}  

In the ideal setting with noiseless two-qubit gates, it has been established that approximate RCS is hard in the average case \cite{Aaronson2016,Bouland2019,Movassagh2018,Movassagh2019}. In particular, the approximate hardness implies that the classical intractability of RCS persists even in the presence of the total variation distance noise given in Eq.\ \eqref{eq:total variation distance noise}. On the other hand, it is important to realize that realistic quantum devices are not able to sample, even approximately, from an ideal output distribution $P_{\mathcal{C}}$ in the limit of large system size. In particular, all the gates in realistic
quantum devices are corrupted by an error channel with a non-zero gate error rate and thus the noisy system eventually converges to a depolarized state. As a result, the fidelity between an actual output state obtained from a noisy quantum circuit $\mathcal{C'}$ and an ideal output state obtained from a noiseless quantum circuit $\mathcal{C}$ decreases exponentially in the system size \cite{Arute2019}. 

For these reasons, the adversarial total variation distance noise model is not directly relevant to realistic settings. Therefore, it is important to investigate noisy versions of RCS where each two-qubit gate is corrupted a noisy two-qubit CPTP map $\mathcal{N}$. Specifically, we numerically study noisy RCS in 1D architecture by using matrix product operators (MPOs). By doing so, we explore the effects of noise on the computational power of a 1D noisy quantum device.

\subsection{Noise model: Two-qubit depolarization channel}
\label{subsection:Noise model}

To make the discussion concrete, we assume that the noise map $\mathcal{N}$ is given by a two-qubit depolarization channel $\mathcal{N}_{2}[p]$ with an error rate $p$. The two-qubit depolarization channel $\mathcal{N}_{2}[p]$ is defined as 
\begin{align}
\mathcal{N}_{2}[p](\hat{\rho}) \equiv (1-p)\hat{\rho} + \frac{p}{15}  \sum_{\hat{P} \in \mathcal{E}_{2}  } \hat{P} \hat{\rho}\hat{P}, 
\end{align}
where $\mathcal{E}_{2} \equiv  \lbrace \hat{I},\hat{X},\hat{Y},\hat{Z} \rbrace^{\otimes 2} - \lbrace \hat{I}\otimes \hat{I} \rbrace$ is the set of $15$ non-trivial two-qubit Pauli operators and $\lbrace \hat{I},\hat{X},\hat{Y},\hat{Z}\rbrace$ is the set of single-qubit Pauli operators. Thus, we incorporate the possibility of correlated two-qubit errors that occur during a two-qubit gate. Note also that the two-qubit depolarization channel $\mathcal{N}_{2}[p]$ can also be expressed as
\begin{align}
\mathcal{N}_{2}[p](\hat{\rho}) = \Big{(} 1-\frac{16}{15}p \Big{)} \hat{\rho} + \frac{16}{15}p \Big{(} \frac{\hat{I}\otimes \hat{I}}{4} \Big{)} \textrm{Tr}[\hat{\rho}]  , 
\end{align}
where $\hat{I}\otimes \hat{I}/4$ is the maximally mixed (or completely depolarized) two-qubit state. 

In the context of fault-tolerant quantum computing, two-qubit depolarization channels are used to model errors that happen during two-qubit gates, such as CNOT and CZ gates (i.e., to perform a detailed circuit-level noise analysis) \cite{Fowler2012,Fowler2012a,Chamberland2019,Chamberland2020,Chamberland2020a,
Das2020,Delfosse2020,Delfosse2020a,Chamberland2020b}. For instance, the fault-tolerance threshold of the surface-code $p\simeq 0.01$ \cite{Fowler2012} is obtained by applying the two-qubit depolarization channel $\mathcal{N}_{2}[p]$ after each two-qubit gate. Since any two-qubit errors can be converted via a noise twirling \cite{Emerson2007} to a two-qubit depolarization channel, the latter serves as a pessimistic noise model for a detailed circuit-level fault-tolerance anaylsis. 

In the context of noisy RCS, we consider the two-qubit depolarization model simply as a representative error model. We expect that different error models will give rise to the same conclusions we reach with the two-qubit depolarization model. Moreover, the MPO method is completely general and applies to any two-qubit error model.    

\section{Matrix product operators}
\label{section:Matrix product operators}

In this section, we briefly review matrix product operators (MPOs) \cite{Verstraete2004,Zwolak2004} which we use to analyze 1D noisy RCS. We also introduce MPO entanglement entropy that determines the cost of classical simulation based on MPOs. While we focus on qubits in later sections, we consider qudits here to make the review as general as possible.    

\subsection{Vectorization of density matrices and canonical form}
\label{subsection:Vectorization of density matrices}

Consider a general $n$-qudit density matrix 
\begin{align}
\hat{\rho} &= \sum_{i_{1},\cdots,i_{n}, i'_{1},\cdots,i'_{n} =0}^{d-1} \rho_{i_{1}i'_{1},\cdots,i_{n}i'_{n}} |i_{1}\cdots i_{n}\rangle\langle i'_{1}\cdots i'_{n}|, 
\end{align} 
where $\rho_{i_{1}i'_{1},\cdots,i_{n}i'_{n}} \equiv \langle i_{1}\cdots i_{n}| \hat{\rho} | i'_{1}\cdots i'_{n} \rangle$. We first vectorize the density matrix by mapping the bra $\langle i'_{l}|$ to a ket $|\bar{i}'_{l}\rangle$, i.e., $\hat{\rho}\rightarrow |\hat{\rho}\rrangle$ where 
\begin{align}
|\hat{\rho}\rrangle  &= \sum_{i_{1},\cdots,i_{n}, i'_{1},\cdots,i'_{n} =0}^{d-1}  \rho_{i_{1}i'_{1},\cdots,i_{n}i'_{n}} |i_{1}\bar{i}'_{1}\cdots i_{n}\bar{i}'_{n}\rrangle . 
\end{align}
Then, to make the notation simpler, we merge the indices $i_{l}$ and $\bar{i}'_{l}$ and define $I_{l} = d \cdot i_{l} + \bar{i}'_{l} \in \lbrace 0,\cdots ,d^{2}-1 \rbrace$ to get  
\begin{align}
|\hat{\rho}\rrangle  &= \sum_{I_{1},\cdots, I_{n} =0 }^{d^{2}-1} \rho_{I_{1},\cdots, I_{n}} |I_{1}\cdots I_{n}\rrangle.  
\end{align}   
In MPO representation, we write 
\begin{align}
\rho_{I_{1},\cdots, I_{n}} &= \sum_{\alpha_{1},\cdots, \alpha_{n-1} =0 }^{\chi-1}  A^{[1] I_{1}}_{\alpha_{1}} A^{[2]I_{2}}_{\alpha_{1}\alpha_{2}} \cdots A^{[n]I_{n}}_{\alpha_{n-1}} . 
\end{align}
Here, $\chi$ is called the bond dimension. Note that for even $n$, MPOs with a bond dimension $\chi \ge d^{n}$ can represent any $n$-qudit density matrix $\hat{\rho}$ exactly. 

In the canonical form, a vectorized MPO is given by 
\begin{align}
|\hat{\rho}\rrangle &= \sum_{I_{1},\cdots, I_{n} =0 }^{d^{2}-1} \sum_{\alpha_{1},\cdots,\alpha_{n-1}=0}^{\chi-1} \Gamma^{[1]I_{1}}_{\alpha_{1}} \lambda^{[1]}_{\alpha_{1}} \Gamma^{[2]I_{2}}_{\alpha_{1}\alpha_{2}} \lambda^{[2]}_{\alpha_{2}}
\nonumber\\
&\qquad\qquad\quad \cdots  \lambda^{[n-1]}_{\alpha_{n-1}} \Gamma^{[n]I_{n}}_{\alpha_{n-1}} |I_{1}I_{2}\cdots I_{n}\rrangle, \label{eq:MPO canonical form}
\end{align}
such that 
\begin{align}
|\hat{\rho}\rrangle &= \sum_{\alpha_{l}=0}^{\chi-1}\lambda^{[l]}_{\alpha_{l}} |\Phi^{[1\cdots l]}_{\alpha_{l}}\rrangle |\Phi^{[(l+1)\cdots n]}_{\alpha_{l}}\rrangle 
\end{align}
is a Schmidt decomposition of the vectorized density matrix $|\hat{\rho}\rrangle$ with respect to the cut $[1\cdots l] : [(l+1)\cdots n]$ for all $l \in \lbrace 1,\cdots, n-1 \rbrace$. Also, $\lambda^{[l]}_{\alpha_{l}}$ are called the singular values and
\begin{align}
|\Phi_{\alpha_{l}}^{[1\cdots l]}\rrangle &= \sum_{I_{1},\cdots, I_{l}=0}^{d^{2}-1}\sum_{\alpha_{1},\cdots,\alpha_{l-1}=0}^{\chi-1}  
\nonumber\\
&\qquad \Gamma_{\alpha_{1}}^{ [1] I_{1}} \lambda_{\alpha_{1}}^{[1]} \cdots \Gamma_{\alpha_{l-1}\alpha_{l}}^{ [l] I_{l}} |I_{1} \cdots I_{l}\rrangle, 
\nonumber\\
|\Phi_{\alpha_{l}}^{[(l+1)\cdots n]}\rrangle &= \sum_{I_{l+1},\cdots, I_{n}=0}^{d^{2}-1} \sum_{\alpha_{l+1},\cdots,\alpha_{n-1}=0}^{\chi-1}
\nonumber\\
& \Gamma_{\alpha_{l}\alpha_{l+1}}^{ [l+1] I_{l+1}}  \cdots  \lambda_{\alpha_{l-1}}^{[n-1]} \Gamma_{\alpha_{n-1}}^{ [n] I_{n}} |I_{l+1} \cdots I_{n}\rrangle 
\end{align}
are orthonormalized. That is, we have
\begin{align}
&\llangle \Phi^{[1\cdots l]}_{\alpha_{l}} |\Phi_{\beta_{l}}^{[1\cdots l]}\rrangle = \delta_{\alpha_{l}\beta_{l}}, 
\nonumber\\
&\llangle \Phi^{[(l+1)\cdots n]}_{\alpha_{l}} |\Phi_{\beta_{l}}^{[(l+1)\cdots n]}\rrangle = \delta_{\alpha_{l}\beta_{l}} ,
\end{align}  
for all $l\in \lbrace 1,\cdots, n-1 \rbrace$. 

\subsection{Canonical update of MPOs}

In 1D noisy RCS, we keep applying two-qudit CPTP maps on two neighboring qudits, say the $l^{\textrm{th}}$ and the $l+1^{\textrm{th}}$ qudits. Thus, to simulate 1D noisy RCS using MPOs, it is important to update an MPO in a canonical way after applying each two-qudit CPTP map, so that we are guaranteed to be left with an updated MPO in the canonical form. In the case of matrix product states (MPSs) that are pure states, a canonical update upon the action of a local unitary operator can be done locally \cite{Vidal2003}. For instance, if a single-qudit unitary operation $\hat{U}$ is applied to the $l^{\textrm{th}}$ qudit, one can simply update $\Gamma^{[l]i_{l}}_{\alpha_{l-1}\alpha_{l}}$ as 
\begin{align}
\Gamma^{[l]i_{l}}_{\alpha_{l-1}\alpha_{l}} &\leftarrow \sum_{j_{l}=0}^{d-1} U_{i_{l}j_{l}} \Gamma^{[l]j_{l}}_{\alpha_{l-1}\alpha_{l}}, 
\end{align}
where $U_{i_{l}j_{l}} \equiv \langle i_{l}|\hat{U}|j_{l}\rangle$. All the other parameters need not be updated because single-qudit unitary operations cannot affect the entanglement structure of the chain. Note that we used $i_{l}$ instead of $I_{l} = di_{l} + \bar{i}'_{l}$ since we are dealing with pure states when considering MPSs. Similarly, if a two-qudit unitary operation is applied to the $l^{\textrm{th}}$ and the $l+1^{\textrm{th}}$ qudits, only 
\begin{align}
\lambda^{[l]}_{\alpha}, \,\,\, \Gamma^{[l] i_{l} }_{\alpha_{l-1}\alpha_{l}}, \,\,\, \Gamma^{[l+1] i_{l+1} }_{\alpha_{l}\alpha_{l+1}}
\end{align}       
need to be updated because the unitary operation can only affect the entanglement along the cut $[1\cdots l]: [l+1\cdots n]$. 

In contrast, when we work with MPOs and CPTP maps, canonical update cannot be done locally any more because CPTP maps are generally not unitary. To elaborate more on this, let us get back to the case of pure states and consider an $n$-qubit GHZ state
\begin{align}
|\textrm{GHZ}_{n}\rangle = \frac{1}{\sqrt{2}} ( |0\rangle^{\otimes n} + |1\rangle^{\otimes n} ). 
\end{align}   
Note that in this case, the bond dimension $\chi=2$ suffices (i.e., $\alpha_{l} \in \lbrace 0,1 \rbrace$ for all $l\in \lbrace 1,\cdots, n-1 \rbrace$) and the canonical $\lambda$ parameters (or singular values, defined similarly as in the case of MPOs) are given by 
\begin{align}
\lambda^{[l]}_{0} = \lambda^{[l]}_{1} = \frac{1}{\sqrt{2}}, \,\,\, \textrm{for all}\,\,\, l\in\lbrace  1,\cdots n-1\rbrace. 
\end{align} 
This is because the Schmidt decomposition of the GHZ state is given by 
\begin{align}
|\textrm{GHZ}_{n}\rangle = \frac{1}{\sqrt{2}} |0\rangle^{\otimes l} |0\rangle^{\otimes n-l} + \frac{1}{\sqrt{2}} |1\rangle^{\otimes l} |1\rangle^{\otimes n-l}, 
\end{align}
for any cut $[1\cdots l]:[(l+1)\cdots n]$. Then, as an example of local non-unitary action on the system, let us consider a non-destructive measurement of the $l^{\textrm{th}}$ qubit in the computational basis $\lbrace |0\rangle , |1\rangle \rbrace$. Note that we would get either $|0\rangle$ or $|1\rangle$, each with $50\%$ probability, as a measurement outcome. Then, conditioned on measuring $|0\rangle$ or $|1\rangle$, the $n$-qubit GHZ state collapses to a product state $|0\rangle^{\otimes n}$ or $|1\rangle^{\otimes n}$. Thus in any case, the updated singular values of the post-measurement state are given by 
\begin{align}
\lambda^{[l]}_{0} = 1 \textrm{ and } \lambda^{[l]}_{1} = 0, \,\,\, \textrm{for all}\,\,\, l\in\lbrace  1,\cdots n-1\rbrace. 
\end{align} 
This example clearly illustrates that even a local action can make a global impact on the entanglement structure of the chain when the action is not unitary. Note that in this example, the bond dimension needed to describe the output product output state is given by $\chi=1$, instead of $\chi=2$. Such a reduction in the required bond dimension is thanks to the fact that the non-destructive measurement decreased non-trivial quantum correlations between disjoint subregions in the chain that were present in the initial GHZ state.  

With this observation in mind, let us now consider the case of mixed states (or MPOs) and two-qudit CPTP maps. Recall that in the 1D noisy RCS, we apply noisy Haar-random two-qubit gates that are corrupted by a CPTP noise map $\mathcal{N}$ (e.g., a two-qubit depolarization channel $\mathcal{N}_{2}[p]$). That is, each time we apply a noisy Haar-random two-qubit gate, we apply a two-qubit CPTP map 
\begin{align}
\mathcal{M} = \mathcal{N} \cdot \mathcal{U} , \label{eq:two-qubit CPTP map unitary corrupted by a noise map} 
\end{align}   
where $\mathcal{U}$ is defined as $\mathcal{U}(\hat{\rho}) \equiv \hat{U}\hat{\rho}\hat{U}^{\dagger}$ and $\hat{U}$ is a $4\times 4$ Haar-random unitary operator. 
Since the system starts from a completely uncorrelated product state $|0\rangle^{\otimes n}$, Haar-random unitary operations will initially make the system more correlated in a quantum way, and thus increase the required bond dimension needed to faithfully describe the system. On the other hand, the noise map $\mathcal{N}$ will generally tend to decrease non-trivial quantum correlations between disjoint regions across the entire chain, reducing the required bond dimension. In particular, we can expect that the effects of noise will eventually take over and the initially developed non-trivial quantum correlations will be washed away as the circuit depth increases. All these expected behaviors will be corroborated numerically in Section \ref{section:Main results} (see Figs.\ \ref{fig:MPOEE_p_15_10_06} and \ref{fig:MPOEE_n_16_various_p}). To do so, we need to update MPOs such that they remain in the canonical form after applying each noisy two-qubit gate. How this is done is explained in a great detail in Appendix \ref{appendix:Canonical update of MPOs}.

\subsection{MPO entanglement entropy}
\label{subsection:MPO entanglement entropy}

We show in Appendix \ref{appendix:Time cost of MPO simulation of 1D noisy RCS} that the time cost of MPO simulation of 1D noisy RCS (using the update method described in Appendix \ref{appendix:Canonical update of MPOs}) is given by
\begin{align}
T=\mathcal{O}( n^{2}D \chi^{3} ). 
\end{align} 
See also Note added below for a very important remark on this scaling of the time cost. Here, $n$ is the number of qubits, $D$ is the circuit depth, and $\chi$ is the bond dimension. Thus, the simulation cost is determined by the bond dimension $\chi$ for a given set of the system size $n$ and the circuit depth $D$. A relevant quantity that determines the required bond dimension $\chi$ is the spectrum the singular values $\lambda^{[l]}_{\alpha_{l}}$ of an MPO in the canonical form. In the case of pure states and MPSs, the singular values are directly related to the known entanglement measures of pure states such as the entanglement entropy \cite{Bennett1996}.  

In the case of mixed states and MPOs, although the singular values $\lambda^{[l]}_{\alpha_{l}}$ are not directly related to known entanglement measures for mixed states \cite{Bennett1996a,Terhal2002,Jarkovsky2020}, they can still be used to characterize the degree of quantum and classical correlations between two disjoint regions $[1\cdots l]: [(l+1)\cdots n]$. Specifically, we consider the following quantity which we call MPO entanglement entropy to measure the degree of quantum and classical correlations in an MPO (see also, e.g., Refs.\ \cite{Prosen2007,Prosen2009,Xu2020} for an earlier use of the MPO entanglement entropy): 
\begin{align}
\mathcal{S}_{l}(\hat{\rho}) &\equiv -\sum_{\alpha_{l}=0}^{\chi-1} \frac{ ( \lambda^{[l]}_{\alpha_{l}} )^{2} }{ \sum_{\beta_{l} =0}^{\chi-1} ( \lambda^{[l]}_{\beta_{l}} )^{2} } \log_{2}\Big{(}  \frac{ ( \lambda^{[l]}_{\alpha_{l}} )^{2} }{ \sum_{\beta_{l} =0}^{\chi-1} ( \lambda^{[l]}_{\beta_{l}} )^{2} }  \Big{)} .
\end{align} 
Here, $l\in \lbrace 1,\cdots, n-1 \rbrace$ and $\lambda^{[l]}_{\alpha_{l}}$ are the singular values of an MPO $|\hat{\rho}\rrangle$ in the canonical form. The MPO entanglement entropy is equivalent to the operator entanglement entropy (introduced in Ref.\ \cite{Zanardi2001}) applied to a density matrix $\hat{\rho}$. Note that we used the symbol $\mathcal{S}$ instead of $S$ to distinguish the MPO entanglement entropy from the usual entanglement entropy of a pure state. We also normalized the spectrum of the squared singular values $(\lambda^{[l]}_{\alpha_{l}})^{2}$ so that they sum up to unity. We took the squared singular values because then the MPO entanglement entropy is reduced to twice the usual entanglement entropy in the case of pure states, i.e., 
\begin{align}
\mathcal{S}_{l}(\hat{\rho} = |\psi\rangle\langle \psi| )  = 2S_{l}(|\psi\rangle) = 2S( \hat{\psi}_{[1\cdots l]} ).
\end{align} 
Here, $S(\hat{\rho}) = -\textrm{Tr}[ \hat{\rho}\log_{2}\hat{\rho} ]$ is the von Neumann entropy and $\hat{\psi}_{[1\cdots l]} \equiv \textrm{Tr}_{[(l+1)\cdots n]}[|\psi\rangle \langle \psi|   ]$ is the reduced density matrix of the state $|\psi\rangle$ with respect to the subsystem $[1\cdots l]$. The additional factor of $2$ is due to the fact that density matrices are composed of both kets and bras whereas states vectors are described by kets only. More specifically, the MPO singular values of a pure state are simply given by the products of the two copies of the MPS singular values. We choose not to divide the MPO entanglement entropy by $2$ because this additional factor reflects the fact that simulating density matrices is computationally more costly than simulating state vectors.   

We emphasize that for mixed states, the MPO entanglement entropy $\mathcal{S}_{l}(\hat{\rho})$ does not correlate with the von Neumann entropy of the reduced density matrix $S(\hat{\rho}_{[1\cdots l]})$ where $\hat{\rho}_{[1\cdots l]}\equiv \textrm{Tr}_{[(l+1)\cdots n]}[\hat{\rho}]$. For instance, for the completely and globally depolarized state $\hat{\rho} = \hat{I}^{\otimes n}/2^{n}$, the von Neumann entropy of the reduced density matrix $\hat{\rho}_{[1\cdots l]} = 2^{\otimes l}/2^{l}$ is given by $S(\hat{\rho}_{[1\cdots l]}) = l$ and thus grows extensively with the system size $l$. However, the completely and globally depolarized state does not have any non-trivial quantum nor classical correlations between disjoint subsystems. That is, the extensive von Neumann entropy only quantifies the extensive noise in the depolarized state, not its correlation. On the other hand, the MPO entanglement entropy of the completely and globally depolarized state vanishes for all subsystem size $l \in \lbrace 1,\cdots, n-1 \rbrace $, i.e., $\mathcal{S}_{l}(\hat{\rho} =\hat{I}^{\otimes n}/2^{n} )=0$. In this sense, the MPO entanglement entropy captures the non-trivial quantum and classical correlations present in a mixed state, separating out the effects of noise.   

In addition to quantifying the degree of non-trivial quantum and classical correlation, the MPO entanglement entropy also determines the cost of classical MPO simulation. In particular, we define the maximum MPO entanglement entropy as follows 
\begin{align}
\mathcal{S}_{\textrm{max}}(\hat{\rho}) &\equiv \textrm{max}_{l \in \lbrace 1,\cdots, n-1 \rbrace } \mathcal{S}_{l}(\hat{\rho}). 
\end{align} 
The maximum MPO entanglement entropy can be used to estimate the required bond dimension $\chi$ needed to describe a mixed state $\hat{\rho}$. Specifically, if the bond dimension $\chi$ satisfies 
\begin{align}
\log_{2}\chi \gg  \mathcal{S}_{\textrm{max}}(\hat{\rho}) , \label{eq:condition on bond dimension log}
\end{align}   
taking only the largest $\chi$ singular values and discarding all the smaller ones will have a negligible effect on the accuracy of an MPO simulation. The cost of MPO simulation depends heavily on the required bond dimension $\chi$ (see Appendices \ref{appendix:Canonical update of MPOs} and \ref{appendix:Time cost of MPO simulation of 1D noisy RCS}). Thus, the maximum MPO entanglement entropy can serve as a metric that characterizes the computational power of a 1D noisy system. In Section \ref{section:Main results}, we numerically demonstrate that in the 1D noisy RCS setting, the maximum MPO entanglement entropy is bounded by a constant independent of the system size (or the number of qubits $n$), implying that 1D noisy random circuits can be simulated efficiently using MPOs.      

We remark that the MPO entanglement entropy is not necessarily monotone under local operations and classical communication (LOCC). For instance, starting from a product state $|0\rangle\langle 0|_{A}\otimes |0\rangle\langle 0|_{B}$, $A$ and $B$ can generate  a classically correlated state 
\begin{align}
\hat{\rho} &= \frac{1}{2}|0\rangle\langle 0|_{A} \otimes |0\rangle\langle 0|_{B}  +  \frac{1}{2}|1\rangle\langle 1|_{A} \otimes |1\rangle\langle 1|_{B}
\nonumber\\
&\leftrightarrow |\hat{\rho}\rrangle = \frac{1}{2}|00\rrangle_{A}\otimes |00\rrangle_{B} + \frac{1}{2}|11\rrangle_{A}\otimes |11\rrangle_{B}, 
\end{align}
via only LOCC. While the initial product state has vanishing MPO entanglement entropy, the output correlated state has $\mathcal{S}(\hat{\rho}) = 1 > 0$. This example thus clearly illustrates that the MPO entanglement entropy does not separate out classical correlations. Consequently, extensive scaling of the MPO entanglement entropy of a quantum circuit does not necessarily imply that circuit is hard to simulate classically. For instance, it is in principle possible that states of a circuit have extensive MPO entanglement entropy due to classical correlations but still are separable and thus easy to be simulated via classical means \cite{Aharonov1996,Harrow2003}. On the other hand, properly bounded MPO entanglement entropy suggests that small MPO bond dimension $\chi$ may suffice and thus efficient simulation is possible by classical MPO simulation. In what follows, we thus focus on the question of whether the MPO entanglement is bounded by a constant independent of the system size. That is, we only aim to address easiness rather than hardness.   

\subsection{Efficient computation of the output probabilities} 
\label{subsection:Efficient computation of the probability of a measurement outcome from an MPO}

Recall that in RCS, all the qubits are measured in the computational basis at the end of the circuit. Here, we explain how the output probabilities can be efficiently computed from an MPO. Given a mixed state $\hat{\rho}$, the probability of getting an output bit string $\vec{x} = ( x_{1},\cdots, x_{n})$ from a computational-basis measurement is given by
\begin{align}
P_{\hat{\rho}}(\vec{x}) &\equiv  \textrm{Tr}\big{[} |\vec{x}\rangle\langle \vec{x} | \hat{\rho} \big{]} . 
\end{align} 
In the vectorized representation, the state $|\vec{x}\rangle\langle \vec{x}|$ is mapped to $|\vec{X}\rrangle$ where $\vec{X} \equiv (d+1)\vec{x} = ((d+1)x_{1},\cdots, (d+1)x_{n})$. Thus, $P_{\hat{\rho}}(\vec{x})$ is given by 
\begin{align}
P_{\hat{\rho}}(\vec{x}) = \llangle \vec{X} | \hat{\rho}\rrangle. 
\end{align}
Plugging in the canonical MPO representation of a mixed state $\hat{\rho}$ in Eq.\ \eqref{eq:MPO canonical form}, we find 
\begin{align}
P_{\hat{\rho}}(\vec{x}) &= \sum_{\alpha_{1},\cdots,\alpha_{n-1}=0}^{\chi-1} \Gamma^{[1] (d+1)x_{1} }_{\alpha_{1}} \lambda^{[1]}_{\alpha_{1}} \Gamma^{[2] (d+1)x_{2}  }_{\alpha_{1}\alpha_{2}} \lambda^{[2]}_{\alpha_{2}} 
\nonumber\\
&\qquad\qquad\qquad\qquad \cdots   \lambda^{[n-1]}_{\alpha_{n-1}} \Gamma^{[n] (d+1)x_{n} }_{\alpha_{n-1}}  . 
\end{align}
Note that this quantity can be efficiently computed via a sequential matrix multiplications of one $1\times \chi$ matrix,  $2n-3$ $\chi\times \chi$ matrices, and one $\chi\times 1$ matrix. Also, we can similarly compute the total probability $\textrm{Tr}[\hat{\rho}]$ as follows: 
\begin{align}
\textrm{Tr}[\hat{\rho}] &= \sum_{\vec{x} \in \lbrace 0,\cdots, d-1 \rbrace^{n} }P_{\hat{\rho}}(\vec{x})
\nonumber\\
&=  \sum_{\alpha_{1},\cdots,\alpha_{n-1}=0}^{\chi-1} \Big{[} \sum_{x_{1}=0}^{d-1} \Gamma^{[1] (d+1)x_{1} }_{\alpha_{1}} \Big{]} \lambda^{[1]}_{\alpha_{1}} 
\nonumber\\
&\qquad\qquad\qquad \times \Big{[} \sum_{x_{2}=0}^{d-1} \Gamma^{[2] (d+1)x_{2}  }_{\alpha_{1}\alpha_{2}} \Big{]} \lambda^{[2]}_{\alpha_{2}}   
\nonumber\\
&\qquad\qquad \cdots   \lambda^{[n-1]}_{\alpha_{n-1}} \Big{[} \sum_{x_{n}=0 }^{d-1}  \Gamma^{[n] (d+1)x_{n} }_{\alpha_{n-1}}  \Big{]}. 
\end{align}
Ideally, the total probability $\textrm{Tr}[\hat{\rho}]$ should be unity but it will be smaller due to the discarded small singular values. In our numerical demonstration below, we compute the total probability $\textrm{Tr}[\hat{\rho}]$ to characterize the truncation errors. 

Sampling from the output probability distribution can also be done efficiently. To do so, we first derive the marginal probability distribution of the first dit $x_{1}$ by computing
\begin{align}
P_{\hat{\rho}}^{[1]}(x_{1}) &= \sum_{\alpha_{1},\cdots,\alpha_{n-1}=0}^{\chi-1} \Gamma^{[1] (d+1)x_{1} }_{\alpha_{1}} \lambda^{[1]}_{\alpha_{1}} 
\nonumber\\
&\qquad\qquad \times  \Big{[} \sum_{x_{2}=0}^{d-1} \Gamma^{[2] (d+1)x_{2}  }_{\alpha_{1}\alpha_{2}} \Big{]} \lambda^{[2]}_{\alpha_{2}}  
\nonumber\\
&\qquad\quad  \cdots   \lambda^{[n-1]}_{\alpha_{n-1}} \Big{[} \sum_{x_{n}=0 }^{d-1}  \Gamma^{[n] (d+1)x_{n} }_{\alpha_{n-1}}  \Big{]}
\end{align} 
for $x_{1}\in\lbrace 0,\cdots, d-1 \rbrace$ and get a sample $\bar{x}_{1}$ from the marginal distribution $P_{\hat{\rho}}^{[1]}(x_{1})$. Then, we compute the conditional probability given the sample $\bar{x}_{1}$, i.e., 
\begin{align}
P_{\hat{\rho}}^{[2|1]}(x_{2} | \bar{x}_{1}) &= \frac{ P_{\hat{\rho}}^{[1,2]}( \bar{x}_{1} ,x_{2} ) }{ P_{\hat{\rho}}^{[1]}( \bar{x}_{1} ) }, 
\end{align} 
where $P_{\hat{\rho}}^{[1,2]}( \bar{x}_{1} ,x_{2} )$ is given by
\begin{align}
P_{\hat{\rho}}^{[1,2]}( \bar{x}_{1} ,x_{2} ) &= \sum_{\alpha_{1},\cdots,\alpha_{n-1}=0}^{\chi-1} \Gamma^{[1] (d+1)\bar{x}_{1} }_{\alpha_{1}} \lambda^{[1]}_{\alpha_{1}}  
\nonumber\\
&\qquad\qquad\qquad \times  \Gamma^{[2] (d+1)x_{2}  }_{\alpha_{1}\alpha_{2}}  \lambda^{[2]}_{\alpha_{2}}   
\nonumber\\
&\quad \cdots   \lambda^{[n-1]}_{\alpha_{n-1}} \Big{[} \sum_{x_{n}=0 }^{d-1}  \Gamma^{[n] (d+1)x_{n} }_{\alpha_{n-1}}  \Big{]}, 
\end{align}  
and then get a sample $\bar{x}_{2}$ from the conditional distribution $P_{\hat{\rho}}^{[2|1]}(x_{2} | \bar{x}_{1})$. One can then sequentially get the remaining sample dits $\bar{x}_{3},\cdots, \bar{x}_{n}$ from the conditional probabilities $P_{\hat{\rho}}^{[3|1,2]}(x_{3} | \bar{x}_{1},\bar{x}_{2})$, $\cdots$, $P_{\hat{\rho}}^{[n|1\cdots (n-1)]}(x_{n} | \bar{x}_{1},\cdots, \bar{x}_{n-1})$ which can be computed similarly as above.    
 
\section{Main results} 
\label{section:Main results}

In this section, we present our main results on the MPO simulation of 1D noisy random quantum circuits. Note that in noisy RCS, it suffices to sample an output bit string $\vec{x} = (\bar{x}_{1},\cdots, \bar{x}_{n})$ from the output distribution of a noisy quantum circuit. However, we aim to achieve a strictly more challenging task. That is, we will directly simulate the density matrix of the system $\hat{\rho}$ in real time using MPOs. As detailed in the previous section, sampling from the output distribution can be straightforwardly done if the output state $\hat{\rho}$ is available. 

\subsection{Numerical results}
\label{subsection:Numerical results}

In the numerical simulation of 1D noisy RCS, we consider qubits (i.e., $d=2$) and start from an input product state $|\vec{0}\rangle = |0\rangle^{\otimes n}$. That is, we initialize the MPO parameters as follows: 
\begin{align}
\lambda^{[l]}_{\alpha_{l}} = \begin{cases}
1 & \alpha_{l}=0 \\
0 & \textrm{otherwise}
\end{cases}, 
\end{align}
for all $l\in \lbrace 1,\cdots, n-1 \rbrace$ and 
\begin{align}
\Gamma^{[1]I_{1}}_{\alpha_{1}} &= \begin{cases}
1 & \alpha_{1}=I_{1}=0 \\
0 & \textrm{otherwise}
\end{cases}, 
\nonumber\\
\Gamma^{[2]I_{2}}_{\alpha_{1}\alpha_{2}} &= \begin{cases}
1 & \alpha_{1}=\alpha_{2}=I_{2}=0 \\
0 & \textrm{otherwise}
\end{cases}, 
\nonumber\\
&\,\,\, \vdots 
\nonumber\\
\Gamma^{[n]I_{n}}_{\alpha_{n-1}} &= \begin{cases}
1 & \alpha_{n-1}=I_{n}=0 \\
0 & \textrm{otherwise}
\end{cases}.
\end{align}
Here, $I_{l}\in \lbrace 0,1,2,3 \rbrace$ and $\alpha_{l} \in \lbrace 0,\cdots, \chi-1 \rbrace$. Note that the MPO constructed with the above parameters is in the canonical form since $|0\rangle^{\otimes l} |0\rangle^{\otimes n-l}$ is the Schmidt decomposition of the input product state $|0\rangle^{\otimes n}$ for all $l \in \lbrace 0,\cdots, n-1\rbrace$. 

In the first time step (or circuit depth $1$), we generate $n/2$ Haar-random two-qubit unitary operators 
\begin{align}
\hat{U}^{[1,2]}_{1},\cdots, \hat{U}^{[n-1,n]}_{1} , 
\end{align}
by sampling $n/2$ Haar-random unitary matrices of size $4\times 4$. Then, we construct corresponding $n/2$ two-qubit CPTP maps by applying the two-qubit depolarization channel $\mathcal{N}_{2}[p]$ with a two-qubit gate error rate $p$, i.e., 
\begin{align}
&\mathcal{M}^{[1,2]}_{1} \equiv \mathcal{N}_{2}[p] \cdot \mathcal{U}^{[1,2]}_{1}, \cdots ,  
\nonumber\\
&\mathcal{M}^{[n-1,n]}_{1} \equiv \mathcal{N}_{2}[p] \cdot \mathcal{U}^{[n-1,n]}_{1}, 
\end{align} 
where $\mathcal{U}^{[1,2]}_{1},\cdots, \mathcal{U}^{[n-1,n]}_{1}$ are defined similarly as in the text below Eq.\ \eqref{eq:two-qubit CPTP map unitary corrupted by a noise map}. We sequentially apply these two-qubit CPTP maps and update the MPO parameters in a canonical way as prescribed in Appendix \ref{appendix:Canonical update of MPOs}. After applying all $n/2$ two-qubit CPTP maps, we compute the MPO entanglement entropies $\mathcal{S}_{l}(|\hat{\rho}\rrangle)$ for all $l\in\lbrace 1,\cdots, n-1\rbrace$ and save them. 

Similarly in the second time step (or circuit depth $2$), we generate $(n-2)/2$ Haar-random two-qubit unitary operators 
\begin{align}
\hat{U}^{[2,3]}_{2},\cdots, \hat{U}^{[n-2,n-1]}_{2} , 
\end{align}
and sequentially apply the corresponding CPTP maps $\mathcal{M}^{[2,3]}_{2},\cdots, \mathcal{M}^{[n-2,n-1]}_{2}$ and update the MPO parameters as described in Appendix \ref{appendix:Canonical update of MPOs}. Then, we compute and save the MPO entanglement entropy $\mathcal{S}_{l}(|\hat{\rho}\rrangle)$ for all $l\in\lbrace 1,\cdots, n-1\rbrace$. 

We keep alternating between these two procedures and save the MPO entanglement entropies at the end of each time step. Since all the two-qubit unitary operators are chosen Haar-randomly, the obtained MPO entanglement entropies will vary across different circuit realizations. Thus, we repeat this entire procedure $N_{s}$ times and take the average of the obtained entanglement entropies over $N_{s}$ circuit realizations. Specifically, at each circuit depth and length of the subsystem $l$, we first average the MPO entanglement entropy $\mathcal{S}_{l}(\hat{\rho})$ over $N_{s}$ circuit realizations. Then, we maximize the averaged MPO entanglement entropy over $l$ to compute the maximum MPO entanglement entropy $\mathcal{S}_{\textrm{max}}(\hat{\rho})$. In all the numerical simulations presented below, we choose $N_{s}=24$.

In Fig.\ \ref{fig:MPOEE_n_8}, we consider the cases with $n=8$ qubits subject to various two-qubit gate error rates $0\le p\le 0.1$. In particular, we choose the bond dimension to be $\chi = 2^{8}=256$ so that there are no errors in the MPO simulation due to the bond dimension truncation. In the case of noiseless two-qubit gates (i.e., $p=0$), the maximum MPO entanglement entropy is achieved in the middle of the chain (i.e., along the cut $[1\cdots 4]:[5\cdots 8]$ or $l=4$) in the large circuit depth limit and converges to 
\begin{align}
\mathcal{S}_{\textrm{max}} = 6.56 . 
\end{align}
Note that this value is the same as twice the average entanglement entropy of the $8$-qubit Haar-random states along the cut $[1\cdots 4]:[5\cdots 8]$, which is given by
\begin{align}
S_{16,16} = \Big{[} \Big{(}\sum_{k=17}^{256}\frac{1}{k}\Big{)} - \frac{15}{32}   \Big{]}\log_{2}e = 3.28\cdots. 
\end{align}  
Here, we used the formula established in Refs.\ \cite{Page1993,Foong1994,Sanchez1995,Sen1996}, i.e., 
\begin{align}
S_{m,n} = \Big{[} \Big{(}\sum_{k=n+1}^{mn}\frac{1}{k}\Big{)} - \frac{m-1}{2n} \Big{]}\log_{2}e .
\end{align} 
and plugged in $m=n=2^{4}$ because each subsystem ($[1\cdots 4]$ and $[5\cdots 8]$) consists of $4$ qubits and thus $16$ states. Such an agreement is consistent with the expectation that the system converges to a Haar-random state in the large circuit depth limit if all the two-qubit gates are noiseless.

\begin{figure}[t!]
\centering
\includegraphics[width=0.45\textwidth]{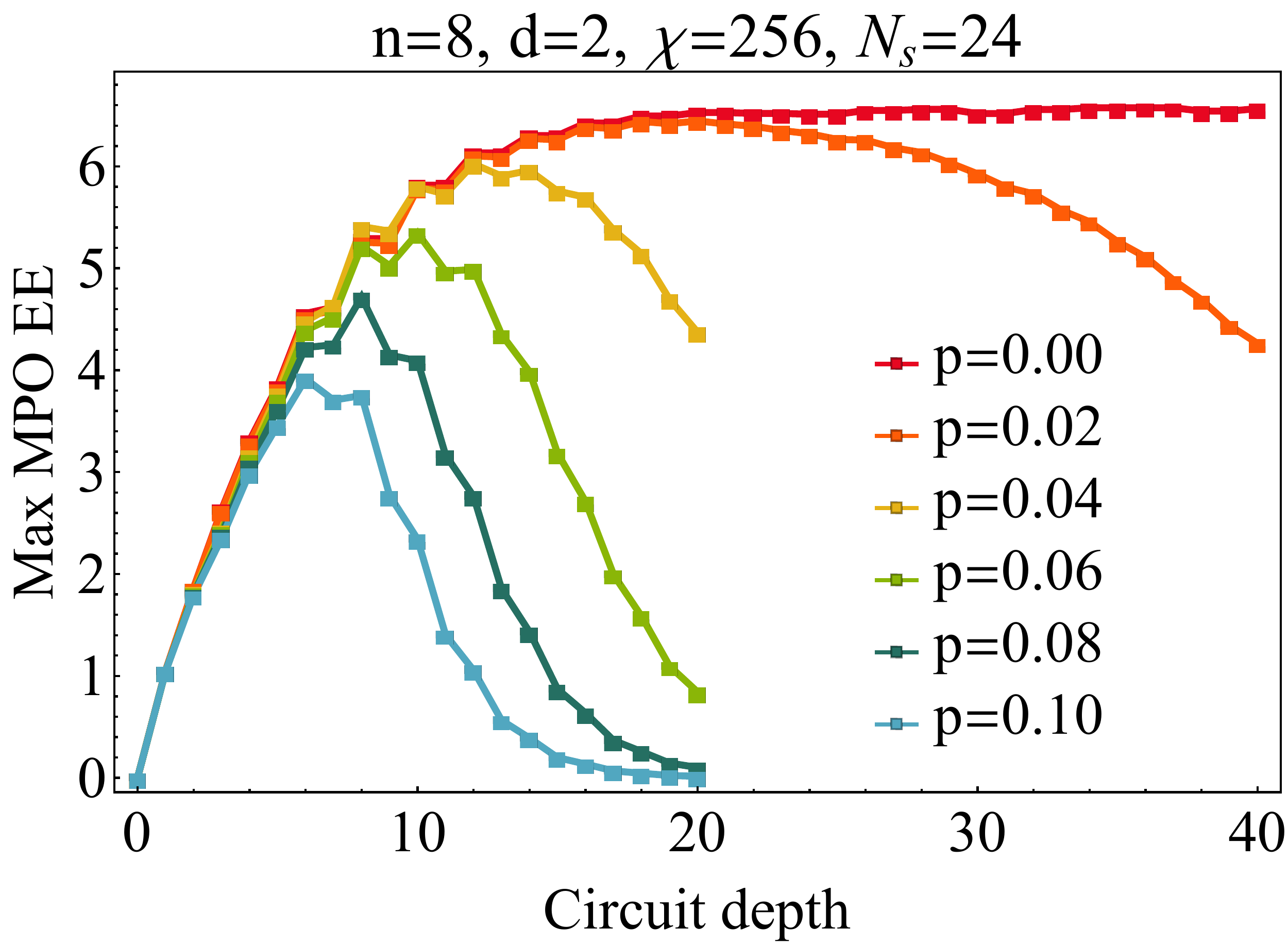}
\caption{Maximum MPO entanglement entropy $\mathcal{S}_{\textrm{max}}$ (averaged over $N_{s}=24$ circuit realizations) as a function of the circuit depth $D$. We took $n=8$ qubits and considered various two-qubit gate error rates $ 0\le p \le 0.1$. The bond dimension was chosen to be $\chi=2^{8} = 256$ so that the MPO simulation is exact. See also Fig.\ \ref{fig:Output_distribution_n_8}.  }
\label{fig:MPOEE_n_8}
\end{figure}

On the other hand, if the two-qubit gates are noisy (i.e., $p\neq 0$), the system eventually converges to the completely and globally depolarized state $\hat{I}^{\otimes 8} / 2^{8}$. As can be seen from Fig.\ \ref{fig:MPOEE_n_8}, for any $p\neq 0$, the maximum MPO entanglement entropy indeed decreases exponentially as the circuit depth increases. This is consistent with the fact that the system converges to the completely and globally depolarized state, which does not possess any non-trivial quantum nor classical correlations. We remark that a similar behavior as in Fig.\ \ref{fig:MPOEE_n_8} was observed in Ref.\ \cite{Dang2017}.

\begin{figure}[t!]
\centering
\includegraphics[width=0.46\textwidth]{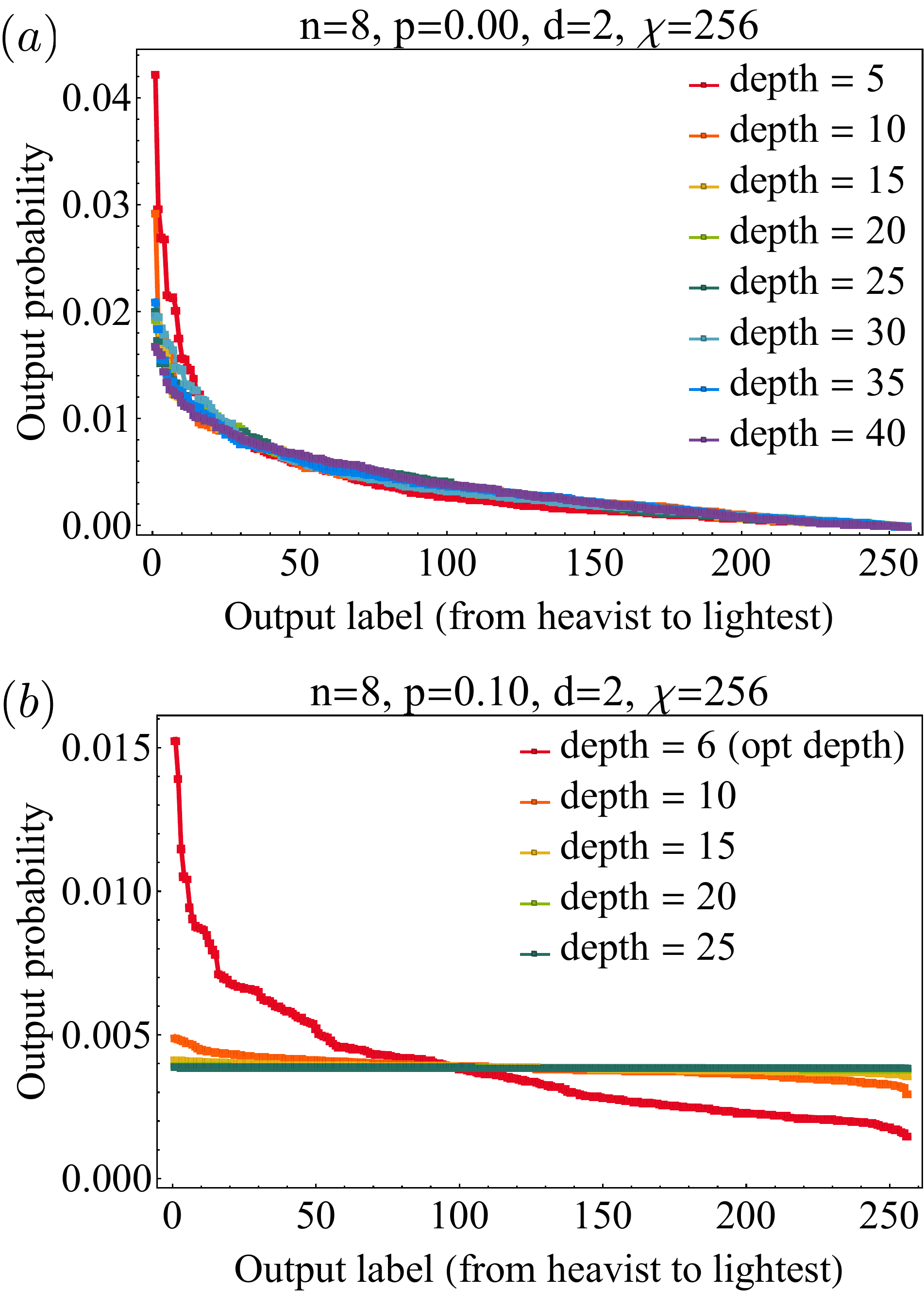}
\caption{Probability of getting an output $n$-bit string $\vec{x} = (x_{1},\cdots, x_{n}) \in \lbrace 0,1\rbrace^{n}$, i.e., $P(\vec{x})$. We took $n=8$ qubits and two-qubit gate error rates (a) $p=0$ and (b) $p=0.1$ and considered various circuit depths for a specific random circuit realization. Note that for $n=8$ qubits, there are $2^{8}=256$ possible output strings $\vec{x}$. We sorted them in a way that a heavier output with larger probability has a smaller label than that of a lighter output. See also Fig.\ \ref{fig:MPOEE_n_8}. To see in what sense the depth $6$ is optimal in the $p=0.1$ case, see Fig.\ \ref{fig:MPOEE_p_15_10_06}(b). }
\label{fig:Output_distribution_n_8}
\end{figure}

\begin{figure}[t!]
\centering
\includegraphics[width=0.48\textwidth]{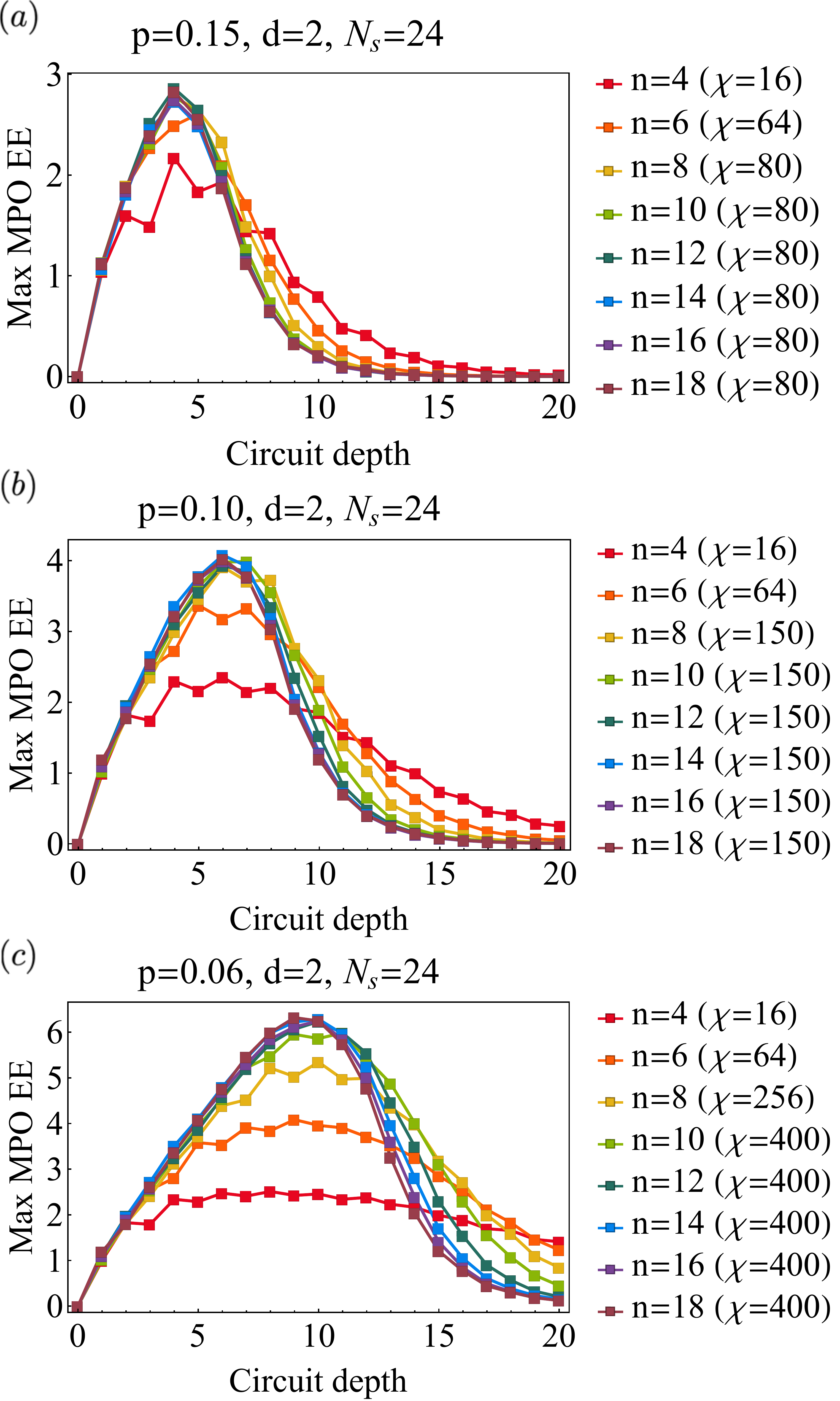}
\caption{Maximum MPO entanglement entropy $\mathcal{S}_{\textrm{max}}$ (averaged over $N_{s}=24$ circuit realizations) as a function of the circuit depth $D$ for various number of qubits $4\le n\le 18$ and two-qubit gate error rates (a) $p=0.15$, (b) $p=0.1$, and (c) $p=0.06$. In all cases, we numerically confirm that the chosen bond dimensions are large enough to account for at least $99.1\%$ of the total probability on average.  }
\label{fig:MPOEE_p_15_10_06}
\end{figure}

In Fig.\ \ref{fig:Output_distribution_n_8}, to get more intuition on what happens in the $n=8$ qubit cases considered in Fig.\ \ref{fig:MPOEE_n_8}, we plot the output probability distribution $P(\vec{x})$ for all possible $2^{8}=256$ outputs $\vec{x} \in \lbrace 0,1 \rbrace^{8}$ for a specific random circuit realization. Note that we ordered the outputs such that a heavier output with larger probability has a smaller label than that of a lighter output. In Fig.\ \ref{fig:Output_distribution_n_8}(a), we consider the noiseless case with the vanishing gate error rate $p=0$. In this case, as discussed above, the system converges to the $8$-qubit Haar-random states in the large circuit depth limit. Correspondingly, the sorted output probability distribution also converges to a fixed distribution in the large circuit depth limit. Note that the output distribution in the large circuit depth limit is far from being uniform. In other words, there are heavier outputs with larger probability that occur more often and lighter outputs that occur less frequently. 

In Fig.\ \ref{fig:Output_distribution_n_8}(b), on the other hand, we consider noisy two-qubit gates with a gate error rate $p=0.1$. In the noisy case, the system eventually converges to the completely and globally depolarized state in the large circuit depth limit. Indeed, as shown in Fig.\ \ref{fig:Output_distribution_n_8}(b), the output probability distribution converges to the trivial uniform distribution in the large circuit depth limit.

In Fig.\ \ref{fig:MPOEE_p_15_10_06}, we consider the cases with a fixed two-qubit gate error rate $p$ and vary the number of qubits $n$. In all the cases we consider, we observe that the MPO entanglement entropy is maximized at a certain optimal circuit depth independent of the system size $n$. Moreover, the maximum achievable MPO entanglement entropy at the optimal circuit depth does not depend on the system size $n$. Consequently, the minimum bond dimension $\chi$ needed to capture the majority of the total probability does not increase exponentially in the number of qubits $n$ in the limit of large $n$.  

For example, when the two-qubit gate error rate is given by $p=0.15$ (see Fig.\ \ref{fig:MPOEE_p_15_10_06}(a)), the MPO entanglement entropy is maximized at an optimal circuit depth $D^{\star}= 4$ for all $n\ge 8$. Also, the maximum achievable MPO entanglement at the optimal circuit depth is given by $\mathcal{S}^{\star}_{\textrm{max}} \simeq 2.75$, which is independent of the system size $n$ as long as $n\ge 8$. Most importantly, we observe that having more than $8$ qubits does not help increasing the MPO entanglement entropy. As a result, the chosen bond dimension $\chi=80$ is at least ten times larger than $2^{\mathcal{S}^{\star}_{\textrm{max}}} \simeq 6.7$ and therefore is large enough to reliably describe the noisy system for any system size $4\le n \le 18$ we considered. Indeed, we numerically confirm that we accounted for at least $99.5\%$ of the total probability (i.e., $\textrm{Tr}[\hat{\rho}] \ge 0.995$) on average with the bond dimension $\chi=80$.          

For a smaller gate error rate $p=0.1$ (see Fig.\ \ref{fig:MPOEE_p_15_10_06}(b)), the optimal circuit depth that maximizes the MPO entanglement entropy is given by $D^{\star} = 6$ for all $n\ge 8$. Also, the corresponding MPO entanglement entropy is given by $\mathcal{S}^{\star}_{\textrm{max}} \simeq 4$ and is independent of the system size above the characteristic system size $n = 8$. The chosen bond dimension $\chi = 150$ is again about ten times larger than $2^{\mathcal{S}^{\star}_{\textrm{max}}} \simeq 16$ and we accounted for at least $99.7\%$ of the total probability on average. Note that higher bond dimension is required in this case than in the case of $p=0.15$, because higher MPO entanglement entropy can be achieved. Note also that the saturation of MPO entanglement entropy occurs with $n=8$ qubits at the optimal circuit depth $D^{\star}=6$. As shown in Fig.\ \ref{fig:Output_distribution_n_8}(b) (see the red line), the sorted output probability distribution at this optimal circuit depth is far from being uniform and thus is still non-trivial. 

For an even smaller gate error rate $p=0.06$ (see Fig.\ \ref{fig:MPOEE_p_15_10_06}(c), we observe that the optimal circuit depth is given by $D^{\star} = 9$ or $D^{\star} = 10$ for all $n\ge 12$ and the maximum achievable MPO entanglement entropy is given by $\mathcal{S}^{\star}_{\textrm{max}} \simeq 6.3$. In this case, the chosen bond dimension $\chi = 400$ is about five times larger than $2^{\mathcal{S}^{\star}_{\textrm{max}}} \simeq 79$ and the accounted total probability is at least $99.1\%$ on average.

\begin{figure}[t!]
\centering
\includegraphics[width=0.47\textwidth]{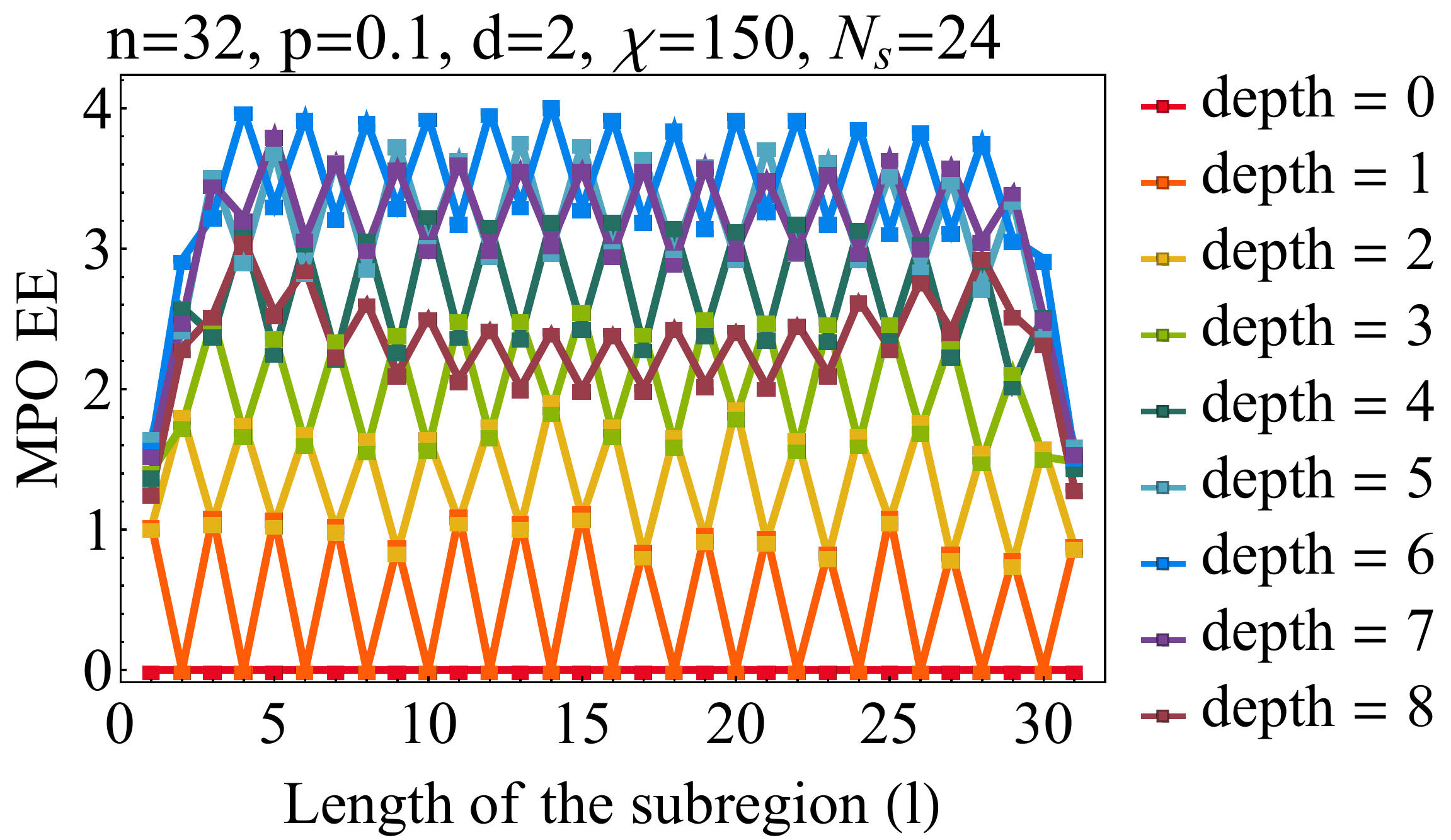}
\caption{MPO entanglement entropy $\mathcal{S}_{l}$ as a function of the length of the subsystem $l$ for various circuit depths. We took $n=32$ qubits and a two-qubit gate error rate $p=0.1$. With the chosen bond dimension $\chi=150$, at least $99.4\%$ of the total probability is accounted for.    }
\label{fig:MPOEE_n_32_p_10_over_length}
\end{figure}

In Fig.\ \ref{fig:MPOEE_n_32_p_10_over_length}, to get more understanding of why the saturation of the MPO entanglement entropy happens, we take the two-qubit gate error rate $p=0.1$ (which was considered in Fig.\ \ref{fig:MPOEE_p_15_10_06}(b)) and consider $n=32$ qubits. We remark that simulating the $32$-qubit system is not so costly because the constant bond dimension $\chi=150$ suffices even for $32$ qubits. In particular, we zoom in to see a more fine-grained MPO entanglement structure and plot the MPO entanglement entropy $\mathcal{S}_{l}$ as a function of the length of the subsystem $l$ (i.e., with respect to the cut $[1\cdots l]:[(l+1)\cdots n]$) for various circuit depths. Note that the largest MPO entanglement entropy is achieved at an optimal circuit depth $D^{\star} =6$, which is consistent with the observation in Fig.\ \ref{fig:MPOEE_p_15_10_06}(b). Notably, we can see that at the optimal circuit depth $D^{\star}=6$ (and far away from the boundaries, i.e., $4\le l \le 28$), the MPO entanglement entropy $\mathcal{S}_{l}$ is independent of the subsystem size $l$. In other words, the MPO entanglement entropy follows an area law at the optimal circuit depth $D^{\star}=6$. This is because the optimal circuit depth is bounded by a constant independent of the system size due to noise and consequently qubits that are not contained within a finite causal cone cannot be correlated with the ones that lie in the causal cone. We provide more discussions on the interplay between noise and the circuit depth in the following section.

\begin{figure}[t!]
\centering
\includegraphics[width=0.47\textwidth]{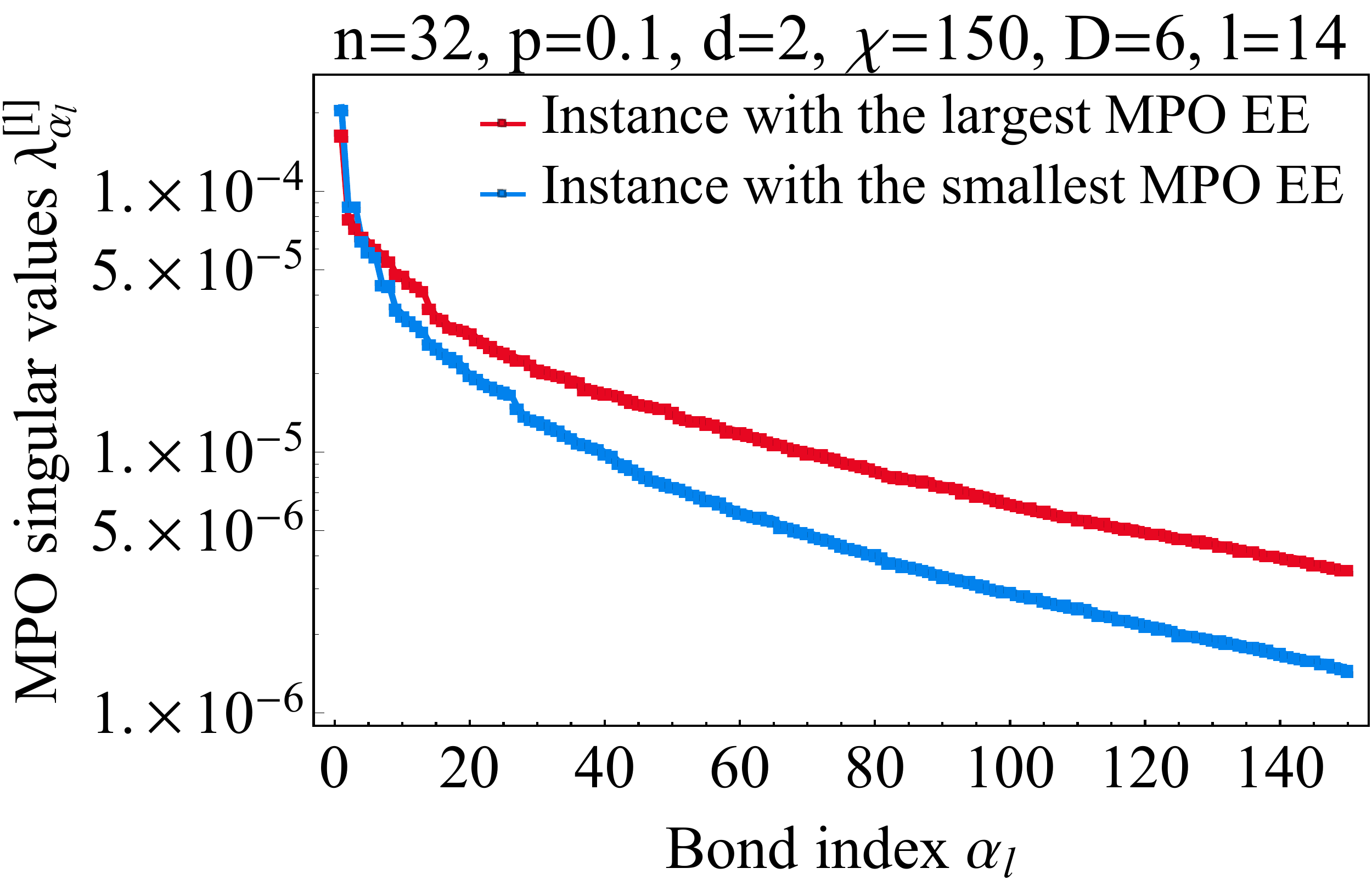}
\caption{MPO singular values $\lambda^{[l]}_{\alpha_{l}}$ for the $n=32$ qubits and the two-qubit gate error rate $p=0.1$ at the optimal circuit depth $D^{\star}=6$. We took the subsystem size $l=14$ where the maximum MPO entanglement entropy is achieved (see Fig.\ \ref{fig:MPOEE_n_32_p_10_over_length}). Among the $N_{s}=24$ circuit instances, we only show two extreme instances with the largest and the smallest MPO entanglement entropy.  }
\label{fig:MPO_spectrum_n_32_p_10_l_14_D_6}
\end{figure}

Lastly in Fig.\ \ref{fig:MPO_spectrum_n_32_p_10_l_14_D_6}, to see the effects of the bond dimension truncation, we plot the spectrum of the MPO singular values $\lambda^{[l]}_{\alpha_{l}}$ for the case with $n=32$ qubit and the two-qubit gate error rate $p=0.1$ considered in Fig.\ \ref{fig:MPOEE_n_32_p_10_over_length}. In particular, we took the optimal circuit depth $D^{\star}=6$ and the subsystem size $l=14$ where the MPO entanglement entropy is maximized. Note that the latter choice is somewhat arbitrary since the MPO entanglement entropy is nearly constant deep inside the bulk (i.e., for $4\le l\le 28$). Among the $N_{s}=24$ circuit realizations, we show the two extreme instances with the largest and the smallest MPO entanglement entropy. In any cases, we can clearly see from Fig.\ \ref{fig:MPO_spectrum_n_32_p_10_l_14_D_6} that the MPO singular values $\lambda^{[l]}_{\alpha_{l}}$ decrease exponentially as the bond index $\alpha_{l}$ increases. Thus, the bond dimension truncation has a negligible effect as long as $\log_{2}\chi$ is much larger than the maximum achievable MPO entanglement entropy $\mathcal{S}^{\star}_{\textrm{max}}$. In particular, due to the exponential decay of the singular values, the required bond dimension $\chi$ as well as the time cost of the MPO simulation would increase only poly-logarithmically in $1/\epsilon$, where $\epsilon$ is the simulation error measured in total variation distance similarly as in Eq.\ \eqref{eq:total variation distance noise} (see, e.g., Ref.\ \cite{Verstraete2006}).

\subsection{Maximum achievable MPO entanglement entropy}
\label{subsection:Maximum MPO entanglement entropy}

Recall that the time cost of MPO simulation of 1D noisy RCS is given by $T = \mathcal{O}(n^{2}D\chi^{3})$ (see Appendix \ref{appendix:Time cost of MPO simulation of 1D noisy RCS}). Thus, the classical MPO simulation cost depends heavily on how large the bond dimension $\chi$ needs to be. As demonstrated above, it suffices in practice to choose the bond dimension $\chi$ such that 
\begin{align}
\chi  = c \cdot  2^{\mathcal{S}^{\star}_{\textrm{max}}}, \label{eq:condition on bond dimension}
\end{align}
for some constant $c\gg 1$ (see also Eq.\ \eqref{eq:condition on bond dimension log}). In Fig.\ \ref{fig:MPOEE_p_15_10_06}, for instance, we chose at least $c\ge 5$ and were able to capture more than $99.1\%$ of the total probability on average. Combining the facts that the simulation cost increases cubically in the bond dimension $\chi$, and the required bond dimension $\chi$ is proportional to $2^{\mathcal{S}^{\star}_{\textrm{max}}}$, we can infer that the simulation cost increases exponentially in the maximum achievable MPO entanglement entropy $\mathcal{S}^{\star}_{\textrm{max}}$. Thus, the maximum achievable MPO entanglement entropy $\mathcal{S}^{\star}_{\textrm{max}}$ can be used as a measure for characterizing the computational power of a 1D noisy quantum system.

\begin{figure}[t!]
\centering
\includegraphics[width=0.47\textwidth]{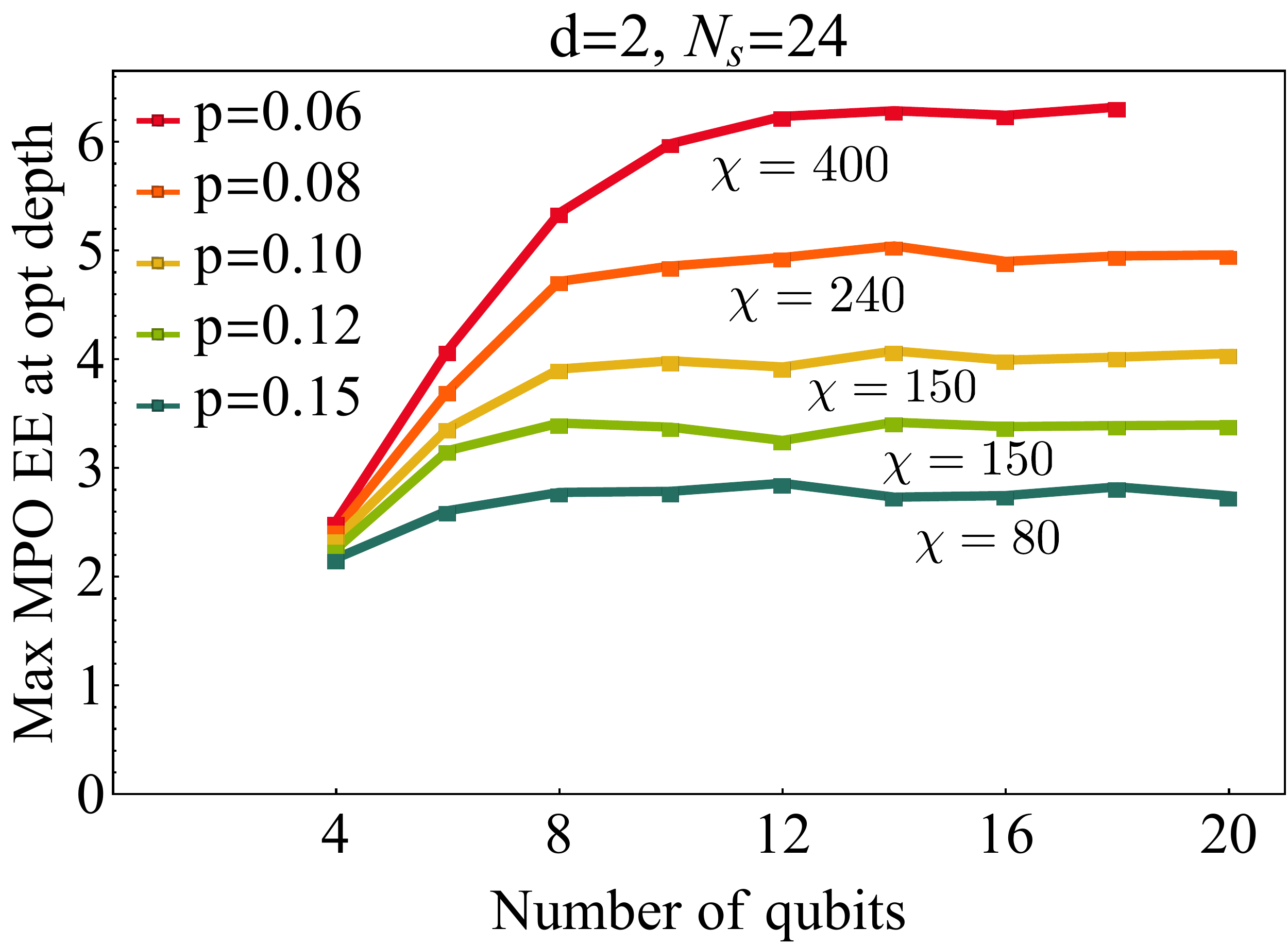}
\caption{Maximum achievable MPO entanglement entropy at the optimal circuit depth $\mathcal{S}^{\star}_{\textrm{max}}$ for various two-qubit gate error rates $0.06\le p\le 0.15$ and number of qubits $4 \le n \le 18$. The bond dimension $\chi$ used in each case is specified next to each curve.  }
\label{fig:MPOEE_Combined}
\end{figure}

In Fig.\ \ref{fig:MPOEE_Combined}, we plot the maximum achievable MPO entanglement entropy $\mathcal{S}^{\star}_{\textrm{max}}$ as a function of the number of qubits $n$ for various two-qubit gate error rates $0.06\le p \le 0.15$. In all the cases, we observe that $\mathcal{S}^{\star}_{\textrm{max}}$ increases linearly in the system size $n$ when the system size is small. In this case, the cost of MPO simulation increases exponentially in the number of qubits $n$. In other words, the computational power of a 1D noisy system increases exponentially in the system size $n$ when $n$ is small. However, as the system size $n$ increases further, the maximum achievable MPO entanglement entropy $\mathcal{S}^{\star}_{\textrm{max}}$ converges to a constant value $\mathcal{S}^{\star}_{\textrm{max},\infty}$ which is independent of the system size $n$. For $p=0.06$, for example, $\mathcal{S}^{\star}_{\textrm{max}}$ converges to $\mathcal{S}^{\star}_{\textrm{max},\infty} \simeq 6.3$. In particular, the saturation occurs around the system size $n=12$. This implies that for $p=0.06$, once we have $n=12$ qubits, adding more qubits does not bring about an exponential growth of the computational power because MPO entanglement entropy remains unchanged and thus the required bond dimension $\chi$ does not increase exponentially in the system size. Indeed, a constant bond dimension $\chi=400$ was sufficient to capture the majority (more than $99.1\%$) of the total probability for all $n\le 18$ in the $p=0.06$ case. Consequently, above the characteristic system size $n=12$, the MPO simulation cost increases only quadratically, not exponentially, in the system size $n$ since $T = \mathcal{O}(n^{2}D\chi^{3})$. 

Note that as the two-qubit gate error rate $p$ becomes larger, the saturation of the MPO entanglement entropy happens at a smaller system size and the saturated value $\mathcal{S}^{\star}_{\textrm{max},\infty}$ becomes smaller. For example, when $p=0.1$, the saturation occurs around the system size $n=8$ and the corresponding MPO entanglement entropy is given by $\mathcal{S}^{\star}_{\textrm{max},\infty} \simeq 4$. In this case, the bond dimension $\chi=150$ is sufficient to capture more than $99.7\%$ of the total probability on average for all $n\le 18$. Note that we need $\chi= 2^{8}= 256$ to exactly describe an $8$-qubit system. Since a smaller bond dimension $\chi=150$ suffices for $p=0.1$, we can infer that a 1D noisy system with a two-qubit gate error rate $p=0.1$ cannot fully occupy the entire $8$-qubit Hilbert space (see also Figs.\ \ref{fig:MPOEE_n_8} and \ref{fig:Output_distribution_n_8}). As a result, having more than $8$ qubits does not bring about an exponential growth of the classical simulation cost.

We have demonstrated in Figs.\ \ref{fig:MPOEE_p_15_10_06} and \ref{fig:MPOEE_Combined} that there exists a characteristic system size, determined by the gate error rate $p$, below which the classical MPO simulation cost increases exponentially in the system size $n$, but above which does so only polynomially in $n$. This is due to the saturation of the MPO entanglement entropy in the large system size limit. If the system size is large enough so that the MPO entanglement entropy is saturated, the saturated MPO entanglement entropy depends solely on the gate error rate $p$, i.e., $\mathcal{S}^{\star}_{\textrm{max},\infty} = \mathcal{S}^{\star}_{\textrm{max},\infty} (p)$. Moreover, the latter becomes the key quantity that determines the required bond dimension $\chi$ and consequently the overall classical MPO simulation cost in the saturated regime.

\begin{figure}[t!]
\centering
\includegraphics[width=0.47\textwidth]{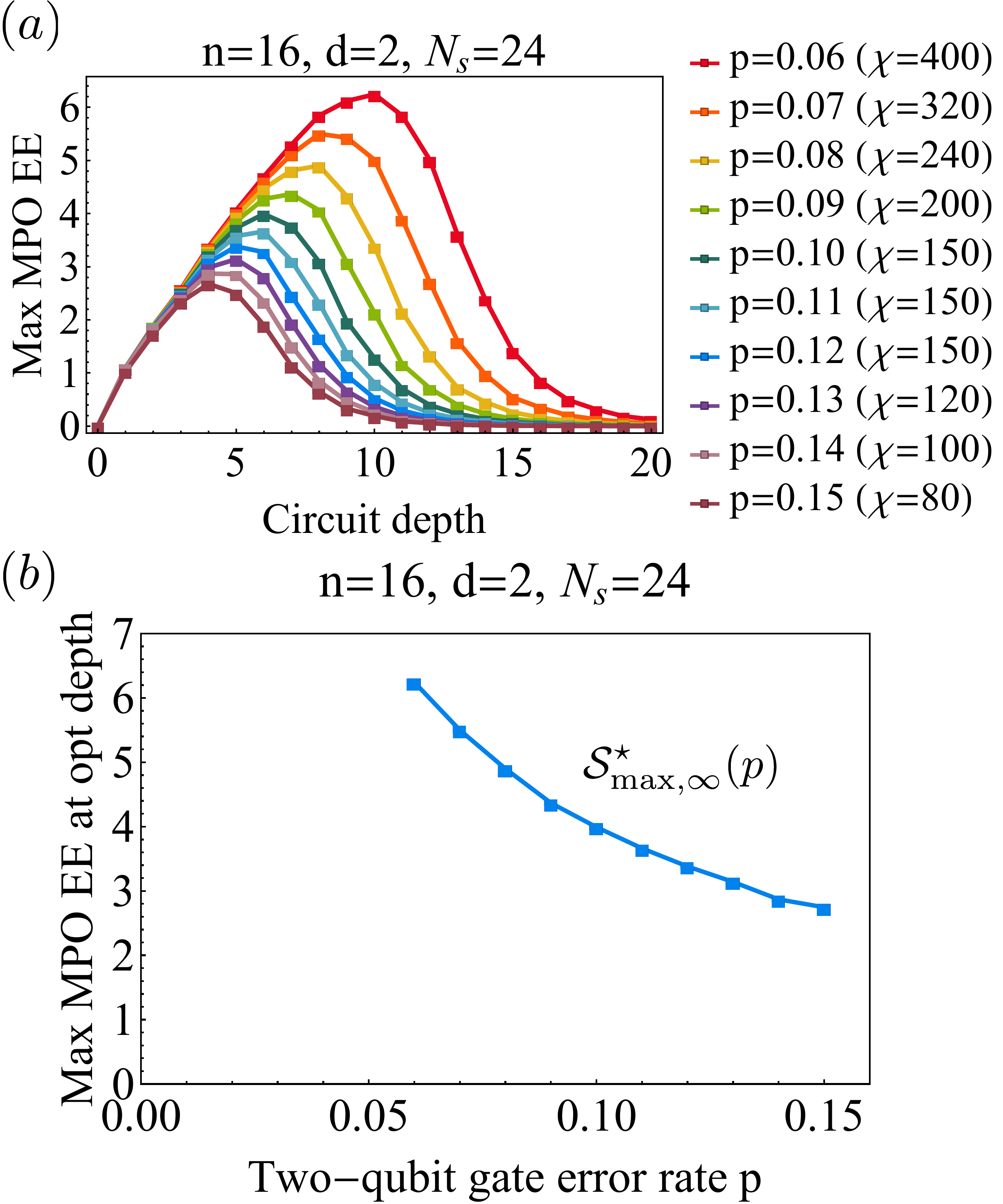}
\caption{(a) Maximum MPO entanglement entropy $\mathcal{S}_{\textrm{max}}$ as a function of the circuit depth $D$ for $n=16$ qubits and various two-qubit gate error rates $0.06\le p \le 0.15$. (b) The saturated values of the maximum achievable MPO entanglement entropy $\mathcal{S}^{\star}_{\textrm{max},\infty}(p)$ as a function of the two-qubit gate error rate $p$ for $0.06 \le p \le 0.15$. To find $\mathcal{S}^{\star}_{\textrm{max},\infty}(p)$, we took the maximum MPO entanglement entropy at the optimal circuit depth $D^{\star}(p)$ for each $p$ from the $n=16$ data shown in (a). Note that for $0.06\le p\le 0.15$, the MPO entanglement entropy is saturated with $n=16$ qubits (see Fig.\ \ref{fig:MPOEE_Combined}).   }
\label{fig:MPOEE_n_16_various_p}
\end{figure}

Note that for the two-qubit gate error rates we considered (i.e., $0.06\le p \le 0.15$), the MPO entanglement entropy is sufficiently saturated with $n=16$ qubits (see Fig.\ \ref{fig:MPOEE_Combined}). In Fig.\ \ref{fig:MPOEE_n_16_various_p}(a), we thus take a closer look into the $n=16$ case and plot the maximum MPO entanglement entropy $\mathcal{S}_{\textrm{max}}$ as a function of the circuit depth $D$ for various two-qubit gate error rates. As can be seen from Fig.\ \ref{fig:MPOEE_n_16_various_p}(a), the optimal circuit depth $D^{\star}(p)$ where the MPO entanglement entropy is maximized increases as the gate error rate $p$ decreases. Moreover, the corresponding MPO entanglement entropy $\mathcal{S}^{\star}_{\textrm{max},\infty}(p)$ increases as $p$ decreases, as shown in Fig.\ \ref{fig:MPOEE_n_16_various_p}(b).

Numerically, we were not able to investigate the cases with low gate error rates (i.e., $p\le 0.06$) because the required bond dimension becomes larger than $\chi=400$ and thus the computing time gets too large. For comparison, note that with the bond dimension $\chi=256$ ($\chi=512$), we can perform an exact MPS simulation of any pure states of a $16$-qubit ($18$-qubit) system. Thus, instead of exploring the low gate error regime numerically, we provide a heuristic analysis of the optimal circuit depth $D^{\star}(p)$ and the maximum achievable MPO entanglement entropy $\mathcal{S}^{\star}_{\textrm{max},\infty}(p)$ in the small $p$ regime. Note that the MPO entanglement entropy is primarily determined by the qubits and gates that are contained within a causal cone. For a given circuit depth $D$, there are $\mathcal{O}(D)$ qubits and $\mathcal{O}(D^{2})$ noisy two-qubit gates (with an error rate $p$ per gate) that are enclosed in the causal cone in the bulk cases (e.g., $l \in [ n/4 , 3n/4 ]$). Thus, one can think of the following heuristic model for the MPO entanglement entropy $\mathcal{S}_{l}$:  
\begin{align}
\mathcal{S}_{l}(D,p) \propto D (1-p)^{cD^{\alpha} } \xrightarrow{p \ll 1} D e^{ - c p  D^{\alpha } }, \label{eq:heuristic model}
\end{align} 
for some constant $c$ and an exponent $\alpha >0$. The prefactor $D$ is due to the linear growth of the MPO entanglement entropy in the absence of noise (see, e.g., Ref.\ \cite{Vidal2008}). The effects of noise are crudely modeled by the term $(1-p)^{cD^{\alpha}} \xrightarrow{p \ll 1} D e^{ - c p  D^{\alpha } }$. The latter term was motivated by the fact that local errors are turned into a global error within the causal cone via random unitary operations. Specifically, we crudely assumed that with probability (or fidelity) $F \sim (1-p)^{cD^{\alpha}}$, the system is in the desired entangled state with an entanglement entropy proportional to $D$, and with probability $1-F$, in the completely and globally depolarized state (within the causal cone) that does not have any non-trivial correlations.  

If all gate errors contribute equally, the exponent $\alpha$ will be given by $\alpha=2$ because there are $\mathcal{O}(D^{2})$ gates in the causal cone. In reality, however, not all gate errors contribute equally because the ones near the bottom of the causal cone are propagated globally to almost all $\mathcal{O}(D)$ qubits in the causal cone and the ones near the top of the causal cone remain almost local. Nevertheless, $\alpha=2$ may still be the case in the large $D$ limit (or equivalently in the small $p$ limit; see below) because the errors, say, in the bottom $10\%$ of the causal cone (accounting for roughly $0.05D^{2} = \mathcal{O}(D^{2})$ error locations) will be almost fully propagated, decreasing the global fidelity $F$ (within the causal cone) by a factor of $1-p$ per each gate error.


Assuming the heuristic model in Eq.\ \eqref{eq:heuristic model}, the optimal circuit depth $D^{\star}(p)$ for a given error rate $p$ is given by 
\begin{align}
D^{\star}(p) = \Big{(} \frac{1}{ \alpha c  p}\Big{)}^{1/ \alpha } \propto p^{-1/ \alpha }, \label{eq:opt depth}
\end{align} 
because $\partial_{D} \mathcal{S}_{l}(D,p) = (1-\alpha c pD^{\alpha})e^{-cpD^{\alpha}} =0$ implies $\alpha c p D^{\alpha}= 1$. Thus, plugging in Eq.\ \eqref{eq:opt depth} to Eq.\ \eqref{eq:heuristic model}, we find
\begin{align}
\mathcal{S}^{\star}_{\textrm{max},\infty}(p) = \mathcal{S}_{l}(D^{\star},p)  \propto p^{-1/ \alpha}. \label{eq:heuristic scaling}
\end{align}
From the data in Fig.\ \ref{fig:MPOEE_n_16_various_p}, we empirically observe $\mathcal{S}^{\star}_{\textrm{max},\infty}(p) = 0.5 p^{-0.9}$ which corresponds to $\alpha = 1.11$. We note that $\alpha < 1$ is not likely to be the case because for the circuit depth $D = \omega( 1/p )$, every each one of the qubits is affected by the noise channel with high probability. The scaling in Eq.\ \eqref{eq:heuristic scaling} implies that the saturation of the MPO entanglement entropy observed in Fig.\ \ref{fig:MPOEE_Combined} will hold for any non-zero gate error rate $p >0$. Thus, our heuristic analysis suggests that there does not exist a non-trivial threshold value of the gate error rate $p$ below which efficient classical simulation is not possible. This is because the required bond dimension $\chi \gg 2^{\mathcal{S}^{\star}_{\textrm{max}}}$ does not grow exponentially in the system size $n$ since the MPO entanglement entropy is saturated. 

On the other hand, the heuristic analysis also suggests that the bond dimension $\chi$ increases exponentially in $(1/p)^{1/\alpha}$. Hence, the classical MPO simulation cost increases exponentially as the gate error rate $p$ decreases, possibly makes classical simulation practically not feasible. This is precisely the reason why we were not able to explore smaller gate error rates than $p=0.06$ in our numerical simulation. In this regards, we remark that the heuristic analysis works in the regime where the optimal circuit depth $D^{\star}$ is large or equivalently when the gate error rate $p$ is small. On the other hand, the gate error rates we considered $0.06\le p\le 0.15$ are not sufficiently small for us to extract the exponent $\alpha$ in a stable manner. Such a quantitative fitting will be reliable only in the low gate error regime. Numerically investigating the low gate error regime would require more advanced computing resources and techniques. We thus leave it as a future research direction.



\section{Relation to previous results}
\label{section:Relation to previous results}

In this section, we compare our results with the related previous results. In essence, our work differs from the previous ones in that we directly simulate mixed states via MPOs that are corrupted by CPTP noise maps. That is, we take advantage of the fact that noise maps (e.g., depolarization channels) reduce non-trivial quantum correlations between disjoint subsystems, making it possible for us to reliably describe the corrupted mixed states with a bounded bond dimension. Indeed, we numerically demonstrate that for 1D noisy random quantum circuits, the MPO entanglement entropy, which characterizes the non-trivial quantum correlations, is bounded by a constant independent of the system size $n$ (see Fig.\ \ref{fig:MPOEE_Combined}). Moreover, the heuristic analysis given in the previous section suggests that the same qualitative behavior will hold for any non-zero gate error rate $p>0$ (see Eq.\ \eqref{eq:heuristic scaling}). Hence, our work suggests that there does not exist a non-zero threshold value of the gate error rate $p$ below which efficient classical simulation is forbidden. In what follows, we explicitly compare these features with the ones observed in the previous works. 

Recently, there have been numerous studies on 1D random quantum circuits subject to local projective (or weak) measurements \cite{Li2018,Chan2019,Skinner2019,Li2019,Szyniszewski2019,Choi2019,
Gullans2019,Gullans2019a,Zabalo2020,Fan2020,Bao2020,Jian2020,
Bera2020,Li2020}. In this model, local projective (or weak) measurements (which happens with a probability $p_{m}$ at each qubit site in a given time step) reduce non-trivial quantum entanglement between disjoint subsystems. In our model, on the other hand, the same role is played by CPTP noise maps. In both models, random unitary operations tend to increase non-trivial quantum correlations, and thus compete with the correlation-decreasing measurements or CPTP noise maps. In the case of projective (or weak) measurements, the 1D system always remains in a pure state. Also, it has been numerically observed that there is a threshold value of the measurement probability ($p_{m}^{(\textrm{th})} \simeq 0.16$) above which area-law entanglement holds. Hence, above the threshold ($p_{m} > p_{m}^{(\textrm{th})}$), dynamics of the random circuits can be efficiently simulated by using MPSs. However, volume-law entanglement holds below the threshold ($p_{m} < p_{m}^{(\textrm{th})}$). Thus, the system cannot be efficiently simulated by using classical computers unless the random unitary operators have a special structure such as being Clifford \cite{Gottesman1997,Gottesman1998,Aaronson2004} or dual-unitary \cite{Piroli2020} operations which can be efficiently simulated regardless of the entanglement structure. 

In contrast, since we consider CPTP noise maps which are realistic models of noise in NISQ devices, the 1D system is in a mixed state and hence we use MPOs to efficiently simulate the system. In particular, CPTP noise maps completely wash away non-trivial quantum correlations and the system eventually reaches the completely and globally depolarized state $\hat{I}^{\otimes n}/2^{n}$ in the large circuit depth limit. On the other hand, in the case of projective (or weak) measurements, there is always a non-zero constant entanglement that survives in the long time limit even in the area-law phase (except for the extreme case with $p_{m}=1$).       
 
We remark that the projective (or weak) measurement models can be used as a sampling-based quantum-trajectory method to simulate 1D noisy RCS subject to CPTP noise maps. For instance, a single-qubit dephasing channel 
\begin{align}
\mathcal{N}_{D}[p](\hat{\rho}) \equiv (1-p_{D})\hat{\rho} + p_{D}\hat{Z}\hat{\rho}\hat{Z} 
\end{align}  
can be understood as a channel that results from performing a non-destructive measurement in the computational basis (or $\hat{Z}$ basis) with a measurement probability $p_{m} = 2p_{D}$ and then forgetting about the measurement outcome, i.e., 
\begin{align}
\mathcal{N}_{D}[p](\hat{\rho}) &= (1-2p_{D})\hat{\rho} + 2p_{D} \Big{(} \frac{\hat{I}+\hat{Z}}{2} \Big{)} \hat{\rho} \Big{(} \frac{\hat{I}+\hat{Z}}{2} \Big{)}
\nonumber\\ 
&\quad + 2p_{D} \Big{(} \frac{\hat{I}-\hat{Z}}{2} \Big{)} \hat{\rho} \Big{(} \frac{\hat{I}-\hat{Z}}{2} \Big{)}.
\end{align} 
Note that $(\hat{I} + \hat{Z} )/2$ and $(\hat{I} - \hat{Z} )/2$ respectively correspond to the projection operators $|0\rangle\langle 0|$ and $|1\rangle\langle 1|$ that project the system onto a space associated with each measurement basis state. We remark that this method can be efficient only when the dephasing error probability $p_{D}$ is above a certain threshold value $p_{D} > p_{m}^{(\textrm{th})} / 2 \simeq 0.08$. However, we stress that the existence of a non-zero threshold value is specific to this quantum-trajectory method and is not intrinsic to the 1D noisy RCS model itself. Our results indicate that while each quantum trajectory in the measurement model yields a pure state with volume-law entanglement below the threshold ($p_{D} < p_{m}^{(\textrm{th})} / 2 \simeq 0.08$), the mixed state that results from putting together all the pure states in all the trajectories will have an area-law MPO entanglement entropy (see also the beginning of Section VIII in Ref.\ \cite{Aharonov2000} for a related discussion). Hence, our method is effective as it directly simulates the mixed states via MPOs and thus maximally takes advantage of the reduction in quantum correlations due to CPTP noise maps. In contrast, such reduction in quantum correlations at the mixed state level is not exploited in the quantum-trajectory method based on the measurement models. 

In another related recent work \cite{Zhou2020}, an MPS method was used to simulate 1D and 2D random quantum circuits. Our work fundamentally differs from this work because in the latter, gate errors are introduced because of the truncation of small singular values when updating MPSs after each two-qubit gate, not because of the CPTP noise channels such as depolarization channels. In other words, Ref.\ \cite{Zhou2020} aims to approximately simulate an ideal random quantum circuit and is not concerned with what types of errors are introduced in the classical simulation as long as a non-zero fidelity is attained (see also the discussion on the difference between the total variation distance noise and the CPTP noise channels in Section \ref{section:Problem setup}). For instance, this work has demonstrated that depth-$20$ 2D random quantum circuits with $54$ qubits can be efficiently simulated up to a global fidelity $\mathcal{F}\ge 0.002$. However, as was pointed out in Ref.\ \cite{Zhou2020}, such a remarkable performance is specific to the setting that they considered where each two-qubit gate is fixed to be the CZ gate and only single qubit gates are chosen to be Haar-random. That is, at the quantitative level, the excellent performance is attributed to a simple grouping strategy available for CZ gates and it will be more costly to simulate a system of the same size if the two-qubit gates are chosen to be Haar-random. Moreover, the method in Ref.\ \cite{Zhou2020} is shown to be efficient only above a certain non-zero threshold value of the gate error rate, i.e., for $\epsilon > \epsilon_{\infty} \sim 0.01$, where $\epsilon$ is the error rate per gate. 

In contrast, while we restricted ourselves to a 1D setting, we specifically addressed gate errors that are introduced due to a practically relevant CPTP noise map (hence we use MPOs instead of MPSs), not due to the bond-dimension truncation in MPO simulations. Moreover, we considered general Haar-random two-qubit gates as opposed to CZ gates. Also in regard to the bond dimension truncation, we demonstrate that a constant bond dimension suffices and the errors associated with the bond-dimension truncation are insignificant in our case. This is again thanks to the fact that we directly simulate mixed states with MPOs and maximally take advantage of the correlation reduction caused by CPTP noise maps. Moreover, our numerical results and heuristic analysis suggest that there does not exist a non-zero threshold value of the gate error rate.  


Lastly in the context of IQP, Ref.\ \cite{Bremner2017} showed that most IQP circuits can be classically simulated approximately if they are subject to depolarization errors with a non-zero error rate. In particular, the simulation efficiency comes from the fact that the output probability distribution $P(\vec{x})$ of an IQP circuit becomes sparse in the Fourier-transformed basis, since high-order Fourier coefficients are suppressed exponentially due to the depolarization errors. A similar technique was used in the context of RCS in Refs.\ \cite{Yung2017,Gao2018} to claim efficient classical simulability of noisy random quantum circuits (see also Ref.\ \cite{Boixo2017} and the supplementary material of Ref.\ \cite{Arute2019} for a related discussion).  


We remark that the methods in Refs.\ \cite{Yung2017,Gao2018} are applicable only in the large circuit depth limit. In contrast, while our MPO simulation is only applicable to 1D settings, it works for any circuit depth $D$. Most importantly, although our MPO simulation is applicable to deep circuits, we are not primarily concerned with the large circuit depth limit. Instead, as illustrated in Fig.\ \ref{fig:Schematic_big_picture}, we have focused on identifying the optimal circuit depth where the maximum non-trivial quantum correlation is attained and understanding how hard (or easy) it is to simulate such an optimal regime.

\section{Summary and open questions}
\label{section:Summary and open questions}

In this work, we have numerically investigated the computational power of 1D noisy quantum systems by using MPOs. The key observation is that the maximum achievable MPO entanglement entropy $\mathcal{S}^{\star}_{\textrm{max}}$ is bounded by a constant $\mathcal{S}^{\star}_{\textrm{max},\infty}(p)$ that is independent of the system size, but depends only on gate error rates (see Fig.\ \ref{fig:MPOEE_Combined}). Our numerical results thus imply that the classical simulation cost of a 1D noisy system increases exponentially in the system size $n$ only until the system size $n$ reaches a certain characteristic system size. Above the characteristic system size, which is determined solely by the gate error rate $p$, adding more qubits brings about only polynomial increase in the classical simulation cost. We have also provided a heuristic argument which suggests that the maximum achievable MPO entanglement entropy would scale as $\mathcal{S}^{\star}_{\textrm{max},\infty}(p) \propto p^{-1/\alpha}$ in the small $p$ limit for some $\alpha >0$. This scaling relation implies that the classical simulation cost of a 1D noisy system would increases exponentially as the gate error rate $p$ decreases. Thus, decreasing the gate error rate $p$ can make the classical MPO simulation practically impossible, given that the system size $n$ reached a certain characteristic system size. For this reason, we were not able to numerically investigate the low gate error rate regime with $p\le 0.06$.   

An immediate future research direction is thus to numerically explore the low gate error regime with an advanced computing resource and see if the same behavior in Fig.\ \ref{fig:MPOEE_Combined} and the scaling in Eq.\ \eqref{eq:heuristic scaling} hold. Moreover, it would be ideal to make the heuristic analysis in Subsection \ref{subsection:Maximum MPO entanglement entropy} more rigorous. These studies will allow us to understand the scaling of the characteristic system size where the saturation of MPO entanglement entropy occurs as a function of the gate error rate $p$ for the currently available gate error rates $p\sim 10^{-3}-10^{-2}$ and the cost of classically simulating such 1D noisy quantum systems via MPOs.  

Another important open question is whether the same conclusions hold also in the case of 2D noisy RCS, which is more relevant to the currently deployed state-of-the-art superconducting qubit systems. A natural way to extend our results to the 2D cases would be to use the projected entangled pair operators (PEPOs), which are a mixed state generalization of the projected entangled pair states (PEPSs) \cite{Verstraete2004a,Verstraete2004b}. 


Specifically, it will be worth exploring if the states in 2D noisy circuits can be faithfully represented by a PEPO with a constant bond dimension that depends only on the gate error rate $p$. However, unlike in the case of MPO, exactly computing an observable from a PEPO is not feasible because exactly contracting PEPSs is $\#$P complete in the worst case \cite{Schuch2007} and in the average case \cite{Haferkamp2020}. Nevertheless, these hardness results do not immediately rule out the possibility of efficient and approximate simulation of 2D noisy RCS because it may be possible to efficiently contract the output PEPOs approximately.

We remark that the computational complexity of 2D RCS with a constant circuit depth has recently been studied in Ref.\ \cite{Napp2020}. In particular, Ref.\ \cite{Napp2020} suggests that all but superpolynomially small fraction of constant-depth 2D random quantum circuits can be simulated approximately and provides numerical evidence supporting the claim. While the two-qubit gates are assumed to be noiseless in this work, the results of Ref.\ \cite{Napp2020} on constant-depth 2D circuits are very relevant to understanding the complexity of 2D noisy RCS. This is because in the presence noise, maximum non-trivial quantum correlations may be achieved at a constant circuit depth. Moreover, since the simulation algorithms for 2D systems given in Ref.\ \cite{Napp2020} are based on simulating 1D systems using MPSs, it would be interesting to see if one can integrate the methods given in Ref.\ \cite{Napp2020} with the MPO approach presented in our work.   


We finally remark that the efficient approximate simulability of typical 2D noisy random circuits (which is subject to future studies) is not in contradiction with the feasibility of fault-tolerant quantum computing with geometrically local interactions \cite{Gottesman2000}. This is because circuits that are used for fault-tolerant quantum computing are far from being random since they are carefully designed so that gate errors are propagated in a restricted manner. If it turns out that typical instances of noisy 2D RCS with Haar-random two-qubit gates can be efficiently simulated (which is again to be explored), another important related question is whether there exists a special class of robust random circuits that are more feasible than fault-tolerant quantum computing, but are structured enough so that gate errors are propagated in a restricted way and thus are harder to simulate classically.    

\textbf{Note added:} We have recently become aware that update of MPO can be done more efficiently than what we suggested here: one can locally move an MPO's center of orthogonality through QR decomposition to bring the MPO in the mixed canonical form for the subsequent gate site and then locally apply time-evolving block decimation (TEBD) method to truncate the excess bond dimension due to the next gate \cite{Schollwock2011}. The time cost of the latter is given by $T = \mathcal{O}(nD\chi^{3})$, as opposed to $\mathcal{O}(n^2D\chi^{3})$ for our method. Regardless, our main claims are not affected because in both algorithms, the time cost increases only polynomially in the system size $n$, provided that the bond dimension $\chi$ is bounded by a constant. Our source code is available upon request.

\section*{Acknowledgments}
\label{section:Acknowledgments}

We would like to thank Fernando Brandao, John Preskill, Norbert Schuch, Soonwon Choi, Michael Gullans, Aidan Dang, Qian Xu, and Roozbeh Bassirianjahromi for helpful discussions. K.N. acknowledges Sunnie Kim for her help in setting up the computing resources, provided by the University of Chicago Research Computing Center, which were used to perform all the numerical simulations in this work. K.N. acknowledges support through the Korea Foundation for Advanced Studies. L.J. acknowledges support from the ARL-CDQI (W911NF-15-2-0067), ARO (W911NF-18-1-0020, W911NF-18-1-0212), ARO MURI (W911NF-16-1-0349), AFOSR MURI (FA9550-15-1-0015, FA9550-19-1-0399), DOE (DE-SC0019406), NSF (EFMA-1640959, OMA-1936118), and the Packard Foundation (2013-39273). B.F. acknowledges support from AFOSR YIP number FA9550-18-1-0148.

\nocite{apsrev41Control}
\bibliography{Noisy_RCS_Quantum,revtex-custom}
%
%
%
%

\clearpage
\appendix

\section{Canonical update of MPOs under the action of a general two-qudit CPTP map}
\label{appendix:Canonical update of MPOs}

Here, we explain in detail how an MPO can be updated in a canonical way upon the action of a two-qudit CPTP map.

\textbf{Bulk case}: Assume that we have an input MPO in the canonical form as described in Eq.\ \eqref{eq:MPO canonical form}. Then, suppose that a two-qudit CPTP map $\mathcal{M}$ is applied to the $l^{\textrm{th}}$ and the $l+1^{\textrm{th}}$ qudits. Here, we consider the bulk case, i.e., $l\in\lbrace 2,\cdots, n-2\rbrace$. Note that the input MPO can be explicitly expressed as    
\begin{align}
|\hat{\rho}\rrangle &= \sum_{J_{l},J_{l+1}=0}^{d^{2}-1} \sum_{\alpha_{l-1},\alpha_{l},\alpha_{l+1}=0}^{\chi-1} |\Phi^{[1\cdots (l-1)]}_{\alpha_{l-1}}\rrangle  
\nonumber\\
&\quad\times  \lambda^{[l-1]}_{\alpha_{l-1}} \Gamma^{[l]J_{l}}_{\alpha_{l-1}\alpha_{l}} \lambda^{[l]}_{\alpha_{l}} \Gamma^{[l+1]J_{l+1}}_{\alpha_{l}\alpha_{l+1}} \lambda^{[l+1]}_{\alpha_{l+1}}  |J_{l}J_{l+1}\rrangle 
\nonumber\\
&\quad\times  |\Phi^{[(l+2)\cdots n]}_{\alpha_{l+1}}\rrangle . 
\end{align}
Define $\mathcal{M}_{I_{l}I_{l+1},J_{l}J_{l+1}}$ as
\begin{align}
\mathcal{M}_{I_{l}I_{l+1},J_{l}J_{l+1}} &\equiv \langle i_{l}i_{l+1}| \mathcal{N} ( |j_{l}j_{l+1}\rangle \langle j'_{l}j'_{l+1}| ) |i'_{l}i'_{l+1}\rangle,    
\end{align}
where $I_{l} = di_{l} + i'_{l}, \cdots, J_{l+1} = d j_{l+1} + j'_{l+1}$. Then, upon the action of the two-qudit CPTP map $\mathcal{M}$, we have as an output state 
\begin{align}
|\hat{\rho}'\rrangle &=\sum_{I_{l},I_{l+1}=0}^{d^{2}-1} \sum_{\alpha_{l-1},\alpha_{l+1}=0}^{\chi-1}  |\Phi^{[1\cdots (l-1)]}_{\alpha_{l-1}}\rrangle  
\nonumber\\
&\times  B^{[l,l+1]}_{I_{l}\alpha_{l-1},I_{l+1}\alpha_{l+1}} |I_{l}I_{l+1}\rrangle   |\Phi^{[(l+2)\cdots n]}_{\alpha_{l+1}}\rrangle, \label{eq:output state l l+1}
\end{align}
where $B^{[l,l+1]}_{I_{l}\alpha_{l-1},I_{l+1}\alpha_{l+1}}$ is defined as 
\begin{align}
B^{[l,l+1]}_{I_{l}\alpha_{l-1},I_{l+1}\alpha_{l+1}}  &\equiv  \sum_{J_{l},J_{l+1}=0}^{d^{2}-1} \sum_{\alpha_{l}=0}^{\chi-1} \mathcal{N}_{I_{l}I_{l+1},J_{l}J_{l+1}}
\nonumber\\
&\!\!\!\!\!\!\!\!\!\!\!\!\!\!\!\!\times \lambda^{[l-1]}_{\alpha_{l-1}}   \Gamma^{[l]J_{l}}_{\alpha_{l-1}\alpha_{l}} \lambda^{[l]}_{\alpha_{l}} \Gamma^{[l+1]J_{l+1}}_{\alpha_{l}\alpha_{l+1}} \lambda^{[l+1]}_{\alpha_{l+1}}  . 
\end{align}
Applying singular value decomposition (SVD) to $B^{[l,l+1]}$, we find 
\begin{align}
B^{[l,l+1]}_{I_{l}\alpha_{l-1},I_{l+1}\alpha_{l+1}} = \sum_{\beta =0}^{d^{2}\chi-1} \lambda'^{[l]}_{\beta} V^{[l]}_{I_{l}\alpha_{l-1},\beta} W^{[l+1]}_{\beta, I_{l+1}\alpha_{l+1} } , \label{eq:SVD B l l+1}
\end{align}
where $V^{[l]}$ and $W^{[l+1]}$ are unitary matrices. Note that the summation index $\beta$ goes from $0$ to $d^{2}\chi-1$ because $B^{[l,l+1]}$ is a $d^{2}\chi\times d^{2}\chi$ matrix. Plugging in Eq.\ \eqref{eq:SVD B l l+1} to Eq.\ \eqref{eq:output state l l+1}, we get the following Schmidt decomposition of the output state $|\hat{\rho}'\rrangle$ with respect to the cut $[1\cdots l]:[(l+1)\cdots n]$: 
\begin{align}
|\hat{\rho}'\rrangle &= \sum_{\beta =0 }^{d^{2}\chi-1} \lambda'^{[l]}_{\beta} |\Phi'^{[1\cdots l]}_{\beta}\rrangle |\Phi'^{[(l+1)\cdots n]}_{\beta}\rrangle, 
\end{align}
where $|\Phi'^{[1\cdots l]}_{\beta}\rrangle$ and $|\Phi'^{[(l+1)\cdots n]}_{\beta}\rrangle$ are given by 
\begin{align}
&|\Phi'^{[1\cdots l]}_{\beta}\rrangle = \sum_{I_{l}=0}^{d^{2}-1} \sum_{\alpha_{l-1}=0}^{\chi-1} V^{[l]}_{I_{l}\alpha_{l-1},\beta} |\Phi^{[1\cdots (l-1)]}_{\alpha_{l-1}}\rrangle  |I_{l}\rrangle, 
\nonumber\\
&|\Phi'^{[(l+1)\cdots n]}_{\beta}\rrangle = \sum_{I_{l+1}=0}^{d^{2}-1} \sum_{\alpha_{l+1}=0}^{\chi-1} W^{[l+1]}_{\beta, I_{l+1}\alpha_{l+1}} |I_{l+1}\rrangle 
\nonumber\\
&\qquad\qquad\qquad\qquad\qquad\quad \times |\Phi^{[(l+2) \cdots n]}_{\alpha_{l+1}}\rrangle , 
\end{align}
and are orthonormalized since $V^{[l]}$ and $W^{[l+1]}$ are unitary matrices. 

Here, we only take the largest $\chi$ singular values, i.e., $\lambda'^{[l]}_{\beta}$ for $\beta \in \lbrace 0,\cdots, \chi-1 \rbrace$ and discard all the smaller singular values to make the bond dimension bounded by $\chi$. Thus, we update the singular values $\lambda^{[l]}_{\alpha_{l}}$ as follows: 
\begin{align}
\lambda^{[l]}_{\alpha_{l}} &\leftarrow \lambda'^{[l]}_{\alpha_{l}} .
\end{align}
Note that the index $\beta$ is replaced by $\alpha_{l}$. In the case of unitary two-qudit gates, all the other singular values are unchanged and we update $\Gamma^{[l] I_{l}}_{\alpha_{l-1}\alpha_{l}}$ and $\Gamma^{[l+1]I_{l+1}}_{\alpha_{l}\alpha_{l+1}}$ accordingly, leaving all the other $\Gamma$ parameters unchanged \cite{Vidal2003}. However, for general two-qudit CPTP maps, this is not the case any more and we should update the singular values and the $\Gamma$ parameters globally. 

To further update the MPO parameters (on the left hand side), note that 
\begin{align}
|\hat{\rho}'\rrangle &= \sum_{\alpha_{l} =0 }^{\chi-1} \lambda'^{[l]}_{\alpha_{l}} |\Phi'^{[1\cdots l]}_{\alpha_{l}}\rrangle |\Phi'^{[(l+1)\cdots n]}_{\alpha_{l}}\rrangle 
\nonumber\\
&=  \sum_{I_{l}=0}^{d^{2}-1} \sum_{\alpha_{l-1},\alpha_{l}=0}^{\chi-1}  (B_{\leftarrow})^{[l-1,l]}_{\alpha_{l-1},I_{l}\alpha_{l}} |\Phi^{[1\cdots (l-1)]}_{\alpha_{l-1}}\rrangle  
\nonumber\\
&\qquad\qquad\qquad\qquad \times |I_{l}\rrangle |\Phi'^{[(l+1)\cdots n]}_{\alpha_{l}}\rrangle, \label{eq:MPO left update start}
\end{align} 
where $B^{[l-1,l]}_{\alpha_{l-1},I_{l}\alpha_{l}}$ is defined as
\begin{align}
(B_{\leftarrow})^{[l-1,l]}_{\alpha_{l-1},I_{l}\alpha_{l}} &\equiv  V^{[l]}_{I_{l}\alpha_{l-1},\alpha_{l}} \lambda'^{[l]}_{\alpha_{l}} . 
\end{align}
Similarly as above, applying SVD to $B^{[l-1,l]}$, we get
\begin{align}
(B_{\leftarrow})^{[l-1,l]}_{\alpha_{l-1},I_{l}\alpha_{l}} &= \sum_{\beta=0}^{\chi-1} \lambda'^{[l-1]}_{\beta} V^{[l-1]}_{\alpha_{l-1}\beta} W^{[l]}_{\beta,I_{l}\alpha_{l}} , 
\end{align}
where $V^{[l-1]}$ is a $\chi\times \chi$ unitary matrix and $W^{[l]}$ is a $d^{2}\chi\times d^{2}\chi$ unitary matrix. Note that the summation index $\beta$ goes from $0$ to $\chi-1$ since $B^{[l-1,l]}$ is a $\chi \times d^{2}\chi$ matrix. Thus, we find the following Schmidt decomposition of the output state with respect to the cut $[1\cdots (l-1)]:[l\cdots n]$:  
\begin{align}
|\hat{\rho}'\rrangle &= \sum_{\beta=0}^{\chi-1} \lambda'^{[l-1]}_{\beta} |\Phi'^{[1\cdots (l-1)]}_{\beta} \rrangle |\Phi'^{[l\cdots n]}_{\beta}\rrangle , 
\end{align}
where $|\Phi'^{[1\cdots (l-1)]}_{\beta} \rrangle$ and $|\Phi'^{[l\cdots n]}_{\beta}\rrangle$ are given by
\begin{align}
&|\Phi'^{[1\cdots (l-1)]}_{\beta} \rrangle = \sum_{\alpha_{l-1}=0}^{\chi-1} V^{[l-1]}_{\alpha_{l-1}\beta} |\Phi^{[1\cdots (l-1)]}_{\alpha_{l-1}}\rrangle, 
\nonumber\\
&|\Phi'^{[l\cdots n]}_{\beta}\rrangle = \sum_{I_{l}=0}^{d^{2}-1} \sum_{\alpha_{l}=0}^{\chi-1}  W^{[l]}_{\beta,I_{l}\alpha_{l}} |I_{l}\rrangle |\Phi'^{[(l+1)\cdots n]}_{\alpha_{l}}\rrangle . 
\end{align}
Consequently, we update $\Gamma^{[l]I_{l}}_{\alpha_{l-1}\alpha_{l}}$ and $\lambda'^{[l-1]}_{\alpha_{l-1}}$ as follows: 
\begin{align}
\Gamma^{[l]I_{l}}_{\alpha_{l-1}\alpha_{l}} &\leftarrow W^{[l]}_{\alpha_{l-1},I_{l}\alpha_{l}} / \lambda'^{[l]}_{\alpha_{l}}, 
\nonumber\\
\lambda^{[l-1]}_{\alpha_{l-1}} &\leftarrow \lambda'^{[l-1]}_{\alpha_{l-1}} . \label{eq:parameter update l-1 l}
\end{align}
Note that the index $\beta$ is replaced by $\alpha_{l-1}$. 

Carrying on, note that 
\begin{align}
|\hat{\rho}'\rrangle &= \sum_{\alpha_{l-1}=0}^{\chi-1} \lambda'^{[l-1]}_{\alpha_{l-1}} |\Phi'^{[1\cdots (l-1)]}_{\alpha_{l-1}} \rrangle |\Phi'^{[l\cdots n]}_{\alpha_{l-1}}\rrangle
\nonumber\\
&= \sum_{\alpha_{l-2},\alpha_{l-1}=0}^{\chi-1}  \sum_{I_{l-1}=0}^{d^{2}-1} (B_{\leftarrow})^{[l-2,l-1]}_{\alpha_{l-2},I_{l-1}\alpha_{l-1}} 
\nonumber\\
&\qquad\qquad \times |\Phi^{[1\cdots (l-2)]}_{\alpha_{l-2}} \rrangle |I_{l-1}\rrangle |\Phi'^{[l\cdots n]}_{\alpha_{l-1}}\rrangle , 
\end{align}
where $(B_{\leftarrow})^{[l-2,l-1]}_{\alpha_{l-2},I_{l-1}\alpha_{l-1}}$ is defined as
\begin{align}
&(B_{\leftarrow})^{[l-2,l-1]}_{\alpha_{l-2},I_{l-1}\alpha_{l-1}} 
\nonumber\\
&\equiv \sum_{\beta=0}^{\chi-1}\lambda^{[l-2]}_{\alpha_{l-2}} \Gamma^{[l-1]I_{l-1}}_{\alpha_{l-2}\beta} V^{[l-1]}_{\beta\alpha_{l-1}}\lambda'^{[l-1]}_{\alpha_{l-1}}. 
\end{align}
Applying SVD to $(B_{\leftarrow})^{[l-2,l-1]}_{\alpha_{l-2},I_{l-1}\alpha_{l-1}}$, we get
\begin{align}
(B_{\leftarrow})^{[l-2,l-1]}_{\alpha_{l-2},I_{l-1}\alpha_{l-1}} &= \sum_{\beta=0}^{\chi-1}\lambda'^{[l-2]}_{\beta} V^{[l-2]}_{\alpha_{l-2}\beta} W^{[l-1]}_{\beta,I_{l-1}\alpha_{l-1}}. 
\end{align}
Similarly as in Eq.\ \eqref{eq:parameter update l-1 l}, we update $\Gamma_{\alpha_{l-2}\alpha_{l-1}}^{[l-1]I_{l-1}}$ and $\lambda_{\alpha_{l-2}}^{[l-2]}$ as follows: 
\begin{align}
\Gamma_{\alpha_{l-2}\alpha_{l-1}}^{[l-1]I_{l-1}} &\leftarrow W^{[l-1]}_{\alpha_{l-2},I_{l-1}\alpha_{l-1}}/\lambda'^{[l-1]}_{\alpha_{l-1}}, 
\nonumber\\
\lambda_{\alpha_{l-2}}^{[l-2]} &\leftarrow \lambda'^{[l-2]}_{\alpha_{l-2}} . 
\end{align}
Repeating the same procedure, $\Gamma_{\alpha_{l-3}\alpha_{l-2}}^{[l-2]I_{l-2}}, \cdots, \Gamma_{\alpha_{1}\alpha_{2}}^{[2]I_{2}} $ and $\lambda_{\alpha_{l-3}}^{[l-3]},\cdots, \lambda_{\alpha_{1}}^{[1]}$ can be updated in the same way. In the very last step, we should update $\Gamma_{\alpha_{1}}^{[1]I_{1}}$ as follows: 
\begin{align}
\Gamma_{\alpha_{1}}^{[1]I_{1}} &\leftarrow \sum_{\beta=0}^{\chi-1} V^{[1]}_{\beta\alpha_{1}}\Gamma_{\beta}^{[1]I_{1}} . \label{eq:MPO left update end}
\end{align}

The remaining MPO parameters on the right hand side are also updated in the same way starting from 
\begin{align}
|\hat{\rho}'\rrangle &= \sum_{\alpha_{l} =0 }^{\chi-1} \lambda'^{[l]}_{\alpha_{l}} |\Phi'^{[1\cdots l]}_{\alpha_{l}}\rrangle |\Phi'^{[(l+1)\cdots n]}_{\alpha_{l}}\rrangle
\nonumber\\
&= \sum_{\alpha_{l},\alpha_{l+1} =0 }^{\chi-1} \sum_{I_{l+1}=0}^{d^{2}-1} (B_{\rightarrow})^{[l+1,l+2]}_{I_{l+1}\alpha_{l},\alpha_{l+1}} 
\nonumber\\
&\qquad\qquad \times |\Phi'^{[1\cdots l]}_{\alpha_{l}}\rrangle |I_{l+1}\rrangle  |\Phi^{[(l+2)\cdots n]}_{\alpha_{l+1}}\rrangle. \label{eq:MPO right update start}
\end{align}
Here, $(B_{\rightarrow})^{[l+1,l+2]}_{I_{l+1}\alpha_{l},\alpha_{l+1}}$ is defined as 
\begin{align}
(B_{\rightarrow})^{[l+1,l+2]}_{I_{l+1}\alpha_{l},\alpha_{l+1}} &\equiv \lambda'^{[l]}_{\alpha_{l}} W^{[l+1]}_{\alpha_{l},I_{l+1}\alpha_{l+1}}. 
\end{align}
Applying SVD to the $d^{2}\chi\times \chi$ matrix $B^{[l+1,l+2]}$, we get
\begin{align}
(B_{\rightarrow})^{[l+1,l+2]}_{I_{l+1}\alpha_{l},\alpha_{l+1}} &= \sum_{\beta=0}^{\chi-1} \lambda'^{[l+1]}_{\beta} V^{[l+1]}_{I_{l+1}\alpha_{l},\beta} W^{[l+2]}_{\beta \alpha_{l+1}}.
\end{align}
and thus update $\Gamma^{[l+1]I_{l+1}}_{\alpha_{l}\alpha_{l+1}}$ and $\lambda^{[l+1]}_{\alpha_{l+1}}$ as follows: 
\begin{align}
\Gamma^{[l+1]I_{l+1}}_{\alpha_{l}\alpha_{l+1}} &\leftarrow V^{[l+1]}_{I_{l+1}\alpha_{l},\alpha_{l+1}} / \lambda'^{[l]}_{\alpha_{l}}, 
\nonumber\\
\lambda^{[l+1]}_{\alpha_{l+1}} &\leftarrow \lambda'^{[l+1]}_{\alpha_{l+1}} . 
\end{align}

Then, we can write 
\begin{align}
|\hat{\rho}'\rrangle &= \sum_{\alpha_{l+1}=0}^{\chi-1} \lambda'^{[l+1]}_{\alpha_{l+1}} |\Phi'^{[1\cdots (l+1)]}_{\alpha_{l+1}}\rrangle |\Phi'^{[(l+2)\cdots n]}_{\alpha_{l+1}}\rrangle 
\nonumber\\
&= \sum_{I_{l+2}=0}^{d^{2}-1} \sum_{\alpha_{l+1},\alpha_{l+2}=0}^{\chi-1}  |\Phi'^{[1\cdots (l+1)]}_{\alpha_{l+1}}\rrangle  |I_{l+2}\rrangle 
\nonumber\\
&\qquad \times (B_{\rightarrow})^{[l+2,l+3]}_{I_{l+2}\alpha_{l+1},\alpha_{l+2}}  |\Phi^{[(l+2)\cdots n]}_{\alpha_{l+2}}\rrangle , 
\end{align}
where $(B_{\rightarrow})^{[l+2,l+3]}_{I_{l+2}\alpha_{l+1},\alpha_{l+2}}$ is defined as
\begin{align}
&(B_{\rightarrow})^{[l+2,l+3]}_{I_{l+2}\alpha_{l+1},\alpha_{l+2}} 
\nonumber\\
&\equiv \sum_{\beta=0}^{\chi-1} \lambda'^{[l+1]}_{\alpha_{l+1}} W^{[l+2]}_{\alpha_{l+1}\beta}  \Gamma^{[l+2]I_{l+2}}_{\beta\alpha_{l+2}} \lambda^{[l+2]}_{\alpha_{l+2}} . 
\end{align}
Applying SVD to $(B_{\rightarrow})^{[l+2,l+3]}_{I_{l+2}\alpha_{l+1},\alpha_{l+2}}$, we get 
\begin{align}
(B_{\rightarrow})^{[l+2,l+3]}_{I_{l+2}\alpha_{l+1},\alpha_{l+2}} &= \sum_{\beta=0}^{\chi-1} \lambda'^{[l+2]}_{\beta} V^{[l+2]}_{I_{l+2}\alpha_{l+1},\beta} W^{[l+3]}_{\beta\alpha_{l+2}}, 
\end{align}
and thus update $\Gamma^{[l+2]I_{l+2}}_{\alpha_{l+1}\alpha_{l+2}}$ and $\lambda^{[l+2]}_{\alpha_{l+2}}$ as follows: 
\begin{align}
\Gamma^{[l+2]I_{l+2}}_{\alpha_{l+1}\alpha_{l+2}} &\leftarrow V^{[l+2]}_{I_{l+2}\alpha_{l+1},\alpha_{l+2}} / \lambda'^{[l+1]}_{\alpha_{l+1}}, 
\nonumber\\
\lambda^{[l+2]}_{\alpha_{l+2}} &\leftarrow \lambda'^{[l+2]}_{\alpha_{l+2}} . 
\end{align} 
$\Gamma^{[l+3]I_{l+3}}_{\alpha_{l+2}\alpha_{l+3}},\cdots, \Gamma^{[n-1]I_{n-1}}_{\alpha_{n-2}\alpha_{n-1}}$ and $\lambda_{\alpha_{l+3}}^{[l+3]}, \cdots, \lambda_{\alpha_{n-1}}^{[n-1]}$ are updated similarly. In the very last step we update $\Gamma^{[n]I_{n}}_{\alpha_{n-1}}$ as follows: 
\begin{align}
\Gamma^{[n]I_{n}}_{\alpha_{n-1}} &\leftarrow \sum_{\beta=0}^{\chi-1}	W^{[n]}_{\alpha_{n-1}\beta}\Gamma^{[n]I_{n}}_{\beta} . \label{eq:MPO right update end}
\end{align}

\textbf{Left edge case}: Let us now we consider a boundary case with $l=1$. Note that an input MPO in the canonical form can be expressed as 
\begin{align}
|\hat{\rho}\rrangle &= \sum_{\alpha_{1},\alpha_{2}=0}^{\chi-1}\sum_{J_{1},J_{2}=0}^{d^{2}-1} \Gamma^{[1]J_{1}}_{\alpha_{1}} \lambda^{[1]}_{\alpha_{1}}\Gamma^{[2]J_{2}}_{\alpha_{1}\alpha_{2}} \lambda^{[2]}_{\alpha_{2}} 
\nonumber\\
&\qquad\qquad\qquad\quad \times  |J_{1}J_{2}\rrangle |\Phi^{[3 \cdots n]}_{\alpha_{2}}\rrangle . 
\end{align}
Upon the action of a two-qubit CPTP map $\mathcal{M}$, we have 
\begin{align}
|\hat{\rho}'\rrangle &=\sum_{\alpha_{2}=0}^{\chi-1} \sum_{I_{1},I_{2}=0}^{d^{2}-1} B^{[1,2]}_{I_{1},I_{2}\alpha_{2}} |I_{1}I_{2}\rrangle |\Phi^{[3 \cdots n]}_{\alpha_{2}}\rrangle , 
\end{align}
where $B^{[1,2]}_{I_{1},I_{2}\alpha_{2}}$ is defined as 
\begin{align}
B^{[1,2]}_{I_{1},I_{2}\alpha_{2}} &= \sum_{\alpha_{1}=0}^{\chi-1} \sum_{J_{1},J_{2}=0}^{d^{2}-1} \mathcal{M}_{I_{1}I_{2},J_{1}J_{2}} 
\nonumber\\
&\qquad\qquad\quad \times \Gamma^{[1]J_{1}}_{\alpha_{1}} \lambda^{[1]}_{\alpha_{1}}\Gamma^{[2]J_{2}}_{\alpha_{1}\alpha_{2}} \lambda^{[2]}_{\alpha_{2}} . 
\end{align}
Applying SVD to $B^{[1,2]}_{I_{1},I_{2}\alpha_{2}}$, we get 
\begin{align}
B^{[1,2]}_{I_{1},I_{2}\alpha_{2}} &= \sum_{\beta=0}^{\chi-1} \lambda'^{[1]}_{\beta} V^{[1]}_{I_{1}\beta} W^{[2]}_{\beta,I_{2}\alpha_{2}} , 
\end{align}
and thus 
\begin{align}
|\hat{\rho}'\rrangle &= \sum_{\beta=0}^{\chi-1} \lambda'^{[1]}_{\beta} |\Phi'^{[1]}_{\beta}\rrangle |\Phi'^{[2\cdots n]}_{\beta}\rrangle  , 
\end{align}
where $|\Phi'^{[1]}_{\beta}\rrangle$ and $|\Phi'^{[2\cdots n]}_{\beta}\rrangle$ are given by 
\begin{align}
|\Phi'^{[1]}_{\beta}\rrangle &= \sum_{I_{1}=0}^{d^{2}-1} V^{[1]}_{I_{1}\beta} |I_{1}\rrangle, 
\nonumber\\
|\Phi'^{[2\cdots n]}_{\beta}\rrangle  &= \sum_{\alpha_{2}=0}^{\chi-1}\sum_{I_{2}=0}^{d^{2}-1} W^{[2]}_{\beta,I_{2}\alpha_{2}}  |I_{2}\rrangle |\Phi^{[3 \cdots n]}_{\alpha_{2}}\rrangle . 
\end{align}
Thus, we update $\Gamma^{[1]I_{1}}_{\alpha_{1}}$ and $\lambda^{[1]}_{\alpha_{1}}$ as follows: 
\begin{align}
\Gamma^{[1]I_{1}}_{\alpha_{1}} &\leftarrow V^{[1]}_{I_{1}\alpha_{1}},
\nonumber\\
\lambda^{[1]}_{\alpha_{1}} &\leftarrow  \lambda'^{[1]}_{\alpha_{1}} . 
\end{align}
Then, the remaining MPO parameters $\Gamma^{[2]I_{2}}_{\alpha_{1}\alpha_{2}},\cdots, \Gamma^{[n]I_{n}}_{\alpha_{n-1}}$ and $\lambda^{[2]}_{\alpha_{2}},\cdots, \lambda^{[n-1]}_{\alpha_{n-1}}$ can be updated in the same way as in the bulk case, following the procedure described in Eqs. \eqref{eq:MPO right update start}--\eqref{eq:MPO right update end}.

\textbf{Right edge case}: Lastly, let us consider another boundary case with $l=n-1$. Note that an input MPO in the canonical form is given by   
\begin{align}
|\hat{\rho}\rrangle &= \sum_{\alpha_{n-2},\alpha_{n-1}=0}^{\chi-1} \sum_{J_{n-1},J_{n}=0}^{d^{2}-1} \lambda^{[n-2]}_{\alpha_{n-2}}\Gamma^{[n-1]J_{n-1}}_{\alpha_{n-2}\alpha_{n-1}} 
\nonumber\\
&\quad\times  \lambda^{[n-1]}_{\alpha_{n-1}}\Gamma^{[n]J_{n}}_{\alpha_{n-1}}   |\Phi^{[1\cdots (n-2)]}_{\alpha_{n-2}}\rrangle |J_{n-1} J_{n}\rrangle. 
\end{align}
Upon the action of a two-qubit CPTP map $\mathcal{M}$, we have 
\begin{align}
|\hat{\rho}'\rrangle &= \sum_{\alpha_{n-2}=0}^{\chi-1} \sum_{I_{n-1},I_{n}=0}^{d^{2}-1}  B^{[n-1,n]}_{I_{n-1}\alpha_{n-2},I_{n}}  
\nonumber\\
&\qquad\qquad\qquad |\Phi^{[1\cdots (n-2)]}_{\alpha_{n-2}}\rrangle    |I_{n-1} I_{n}\rrangle, 
\end{align}
where $B^{[n-1,n]}_{I_{n-1}\alpha_{n-2},I_{n}}$ is defined as 
\begin{align}
B^{[n-1,n]}_{I_{n-1}\alpha_{n-2},I_{n}} &= \sum_{J_{n-1},J_{n}=0}^{d^{2}-1} \sum_{\alpha_{n-1}=0}^{\chi-1} \mathcal{M}_{I_{n-1}I_{n},J_{n-1}J_{n}}
\nonumber\\
&\!\!\!\! \times \lambda^{[n-2]}_{\alpha_{n-2}}\Gamma^{[n-1]J_{n-1}}_{\alpha_{n-2}\alpha_{n-1}} \lambda^{[n-1]}_{\alpha_{n-1}}\Gamma^{[n]J_{n}}_{\alpha_{n-1}} . 
\end{align}
Applying SVD to $B^{[n-1,n]}_{I_{n-1}\alpha_{n-2},I_{n}}$, we get 
\begin{align}
B^{[n-1,n]}_{I_{n-1}\alpha_{n-2},I_{n}} &= \sum_{\beta=0}^{\chi-1} \lambda'^{[n-1]}_{\beta} V^{[n-1]}_{I_{n-1}\alpha_{n-2},\beta} W^{[n]}_{\beta I_{n}}, 
\end{align}
and thus 
\begin{align}
|\hat{\rho}'\rrangle &= \sum_{\beta=0}^{\chi-1} \lambda'^{[n-1]}_{\beta}|\Phi'^{[1\cdots (n-1)]}_{\beta} \rrangle |\Phi'^{[n]}_{\beta}\rrangle, 
\end{align}
where $|\Phi'^{[1\cdots (n-1)]}_{\beta} \rrangle$ and $|\Phi'^{[n]}_{\beta}\rrangle$ are given by 
\begin{align}
|\Phi'^{[1\cdots (n-1)]}_{\beta} \rrangle &= \sum_{\alpha_{n-2}=0}^{\chi-1} \sum_{I_{n-1}=0}^{d^{2}-1}  V^{[n-1]}_{I_{n-1}\alpha_{n-2},\beta}  
\nonumber\\
&\qquad \qquad\qquad  |\Phi^{[1\cdots (n-2)]}_{\alpha_{n-2}}\rrangle |I_{n-1} \rrangle, 
\nonumber\\
|\Phi'^{[n]}_{\beta}\rrangle &= \sum_{I_{n}=0}^{d^{2}-1}  W^{[n]}_{\beta I_{n}} |I_{n}\rrangle. 
\end{align}
Thus, we update $\Gamma^{[n]I_{n}}_{\alpha_{n-1}}$ and $\lambda^{[n-1]}_{\alpha_{n-1}}$ as follows: 
\begin{align}
\Gamma^{[n]I_{n}}_{\alpha_{n-1}} &\leftarrow W^{[n]}_{\alpha_{n-1} I_{n}}, 
\nonumber\\
\lambda^{[n-1]}_{\alpha_{n-1}} &\leftarrow \lambda'^{[n-1]}_{\alpha_{n-1}} . 
\end{align}
Then, the remaining MPO parameters $\Gamma^{[n-1]I_{n-1}}_{\alpha_{n-2}\alpha_{n-1}},\cdots, \Gamma^{[1]I_{1}}_{\alpha_{1}}$ and $\lambda^{[n-1]}_{\alpha_{n-1}},\cdots, \lambda^{[1]}_{\alpha_{1}}$ can be updated in the same way as in the bulk case, following the procedure described in Eqs. \eqref{eq:MPO left update start}--\eqref{eq:MPO left update end}. 

\section{Time cost of MPO simulation of 1D noisy RCS} 
\label{appendix:Time cost of MPO simulation of 1D noisy RCS}

Time cost of performing the canonical update of an MPO upon the action of a two-qudit CPTP map (as prescribed in Appendix \ref{appendix:Canonical update of MPOs}) is given by $\mathcal{O}( n\chi^{3} )$. Here, $\chi^{3}$ is due to the need to perform SVDs of $\mathcal{O}(\chi) \times \mathcal{O}(\chi)$ matrices. Also, $n$ is due to the fact that we need to perform a global update throughout the entire chain even if the two-qudit CPTP map is local. If the CPTP map were a unitary channel, the global update is not necessary and the factor $n$ will be absent. However, in the case of 1D noisy RCS, such a global update is essential to the required bond dimension so the factor $n$ is present. 

In a depth-$D$ 1D noisy RCS with $n$ qubits, we have $ \mathcal{O}( n D) $ noisy Haar-random two-qubit gates. Thus, we should execute the update module described in Appendix \ref{appendix:Canonical update of MPOs} $\mathcal{O}(nD)$ times. Since the time cost of each update is given by $\mathcal{O}( n\chi^{3} )$ as discussed above, the total time cost of MPO simulation of 1D noisy RCS is given by 
\begin{align}
T = \mathcal{O}( n^{2}D \chi^{3} ). 
\end{align}  
See Note added at the end of the main text for a very important remark on this scaling of the time cost.

\end{document}